\documentclass[lettersize,journal]{IEEEtran}
\IEEEoverridecommandlockouts
\usepackage{amsmath}
\usepackage{diagbox}

\usepackage{cite}
\usepackage{amsmath,amssymb,amsfonts}
\usepackage{algorithm}
\usepackage{algorithmicx}
\usepackage{algpseudocode}  % 提供 \Comment、\For、\Require 等
\usepackage{graphicx}
\usepackage{textcomp}
\usepackage{xcolor}
\usepackage{stfloats}
\usepackage{url}
\usepackage{subfig}
\usepackage{graphicx}
\usepackage{bm}
\usepackage{booktabs}
\usepackage{dsfont}
\usepackage{cite}
\usepackage{color}
\usepackage{makecell}
\usepackage{upgreek}
\usepackage{verbatim}
\def\BibTeX{{\rm B\kern-.05em{\sc i\kern-.025em b}\kern-.08em
		T\kern-.1667em\lower.7ex\hbox{E}\kern-.125emX}}
\usepackage{balance}
\usepackage{multirow}
\usepackage{url}
\usepackage{color}
\usepackage{nomencl}
\usepackage{float}
\usepackage{booktabs}
\def\BibTeX{{\rm B\kern-.05em{\sc i\kern-.025em b}\kern-.08em
    T\kern-.1667em\lower.7ex\hbox{E}\kern-.125emX}}
\begin{document}

\title{WiFo-M$^2$: Empower Wireless Communications With Plug-and-Play Environment Sensing via Foundation Model}
% \author{Haotian Zhang,~\IEEEmembership{Graduate Student Member,~IEEE,} Shijian Gao,~\IEEEmembership{Member,~IEEE,}  Xiang Cheng,~\IEEEmembership{Fellow,~IEEE,} 
\author{Haotian Zhang,~\IEEEmembership{Graduate Student Member,~IEEE,}  Shijian Gao,~\IEEEmembership{Member,~IEEE,} Xiang Cheng,~\IEEEmembership{Fellow,~IEEE}
	\thanks{
	This work was supported in part by the National Natural Science Foundation of China under Grant 62125101 and Grant 62401488, in part by the New Cornerstone Science
		Foundation through the Xplorer Prize, and in part by Guangzhou Municipal Science and Technology Project under Grant 2024D03J0008, and Guangdong Provincial Project under Grant 2023ZT10X009. \textit{(Corresponding authors: Xiang Cheng; Shijian Gao.)}
		
	Haotian Zhang and Xiang Cheng are with the State Key Laboratory of Photonics and Communications, School of Electronics, Peking University, Beijing 100871, P. R. China (e-mail: haotianzhang@stu.pku.edu.cn; xiangcheng@pku.edu.cn).

    Shijian Gao is with the Internet of Things Thrust, The Hong Kong University of Science and Technology (Guangzhou), Guangzhou 511400, P. R. China (e-mail: shijiangao@hkust-gz.edu.cn).

    % Liuqing Yang is with the Intelligent Transportation Thrust and Internet of Things Thrust, The Hong Kong University of Science and Technology (Guangzhou), Guangzhou 510000, P. R. China, and the Department of Electronic and Computer Engineering, The Hong Kong University of Science and Technology, Hong Kong SAR, P. R. China (email: lqyang@ust.hk).

}}

\markboth{Journal of \LaTeX\ Class Files, January~2026}%
{Utilize Plug-and-Play Environment Sensing to Activate Intelligent Wireless via Foundation Model}

\maketitle

\begin{abstract}
% The growing adoption of sensor-rich intelligent systems has boosted the use of multi-modal sensing to improve wireless communications. 
The emerging convergence of next-generation wireless networks and agentic artificial
intelligence (AI) is inspiring a new vision: embodied intelligent network entities utilize environmental sensing to refine their physical-layer (PHY) actions. Despite a growing body of preliminary work, prevailing small and task-specific AI models require extensive manual design of data pre-processing, network architecture, and fine-tuning, leaving them tightly coupled to particular PHY actions, system configurations, and deployment scenarios. To address this, we propose a paradigm shift with WiFo-M$^2$, a foundation model that enables environment sensing to be easily integrated into PHY actions, delivering universal performance gains. To extract generalizable out-of-band (OOB) channel-aware features from environment sensing, we introduce ContraSoM, a contrastive pre-training strategy.  Once pre-trained, WiFo-M$^2$ infers future OOB channel-aware features from historical sensory data and strengthens feature robustness via modality-specific data augmentation. Experiments show that WiFo-M$^2$ improves the performance of a comprehensive suite of fundamental PHY actions, demonstrating strong generalization to unseen scenarios.

\end{abstract}


\vspace{-1.6em}
\section{Introduction}
\label{intro}
The 2030 era of networking will feature hundreds of billions of connected devices, immersive applications, and tight cyber physical loops that demand extreme reliability, ultra low latency, and sustainable efficiency \cite{ruichen_COMST}. Meeting these requirements calls for Agentic AI as a foundational paradigm in which autonomous networked agents perceive, reason, plan, and act \cite{ruichen2}. Today, the application of Agentic AI in wireless networks is rapidly expanding, providing a robust framework for autonomous optimization and self-evolving in increasingly complex and dynamic environments \cite{ruichen2}. Against this backdrop, Cheng {\em et al.} recently introduced Embodied Intelligent Wireless (EIW) \cite{EIW} and envisioned it as the new paradigm for next-generation wireless communications. Under EIW, conventional network nodes such as base stations (BSs) and user equipment (UEs) evolve into intelligent network entities endowed with observation–decision–action capabilities.
% , thereby enabling autonomous self-evolution and robust generalization across diverse scenarios.

Driven by the rapid advancement of embodied intelligence, intelligent embodied network entities are increasingly equipped with comprehensive environmental sensing capabilities~\cite{SoM}. Motivated by this trend, the wireless communications community is actively exploring how these ubiquitous sensing sources can be leveraged to enhance physical layer (PHY)  actions. Here, ``actions'' concretely refers to {\color{black}the execution of specific functional tasks within the transceiver pipeline, such as channel estimation, channel prediction, and beamforming.} To systematically give the design principle and underlying mechanisms of such sensing-refined PHY actions, Cheng {\em et al.} proposed the first framework of this field, Synesthesia of Machines (SoM)~\cite{SoM}. {\color{black}SoM establishes a unified framework for multi-modal sensing-communication integration. It provides the theoretical paradigm for mapping diverse environmental sensory data into the wireless channel domain, which bridges the gap between multi-modal observations and the enhancement of PHY actions.} To date, numerous studies have investigated promising applications, such as using multi-modal sensing to boost channel estimation acquisition performance \cite{CLN, TCCN_CP, TCCN_CE}, to streamline beam selection and training \cite{MMFF, vision-beam1, vision-beam,vision-beamT}, and to optimize precoding \cite{ICC}. 

\begin{figure*}
    \centering
    \includegraphics[width=1\linewidth]{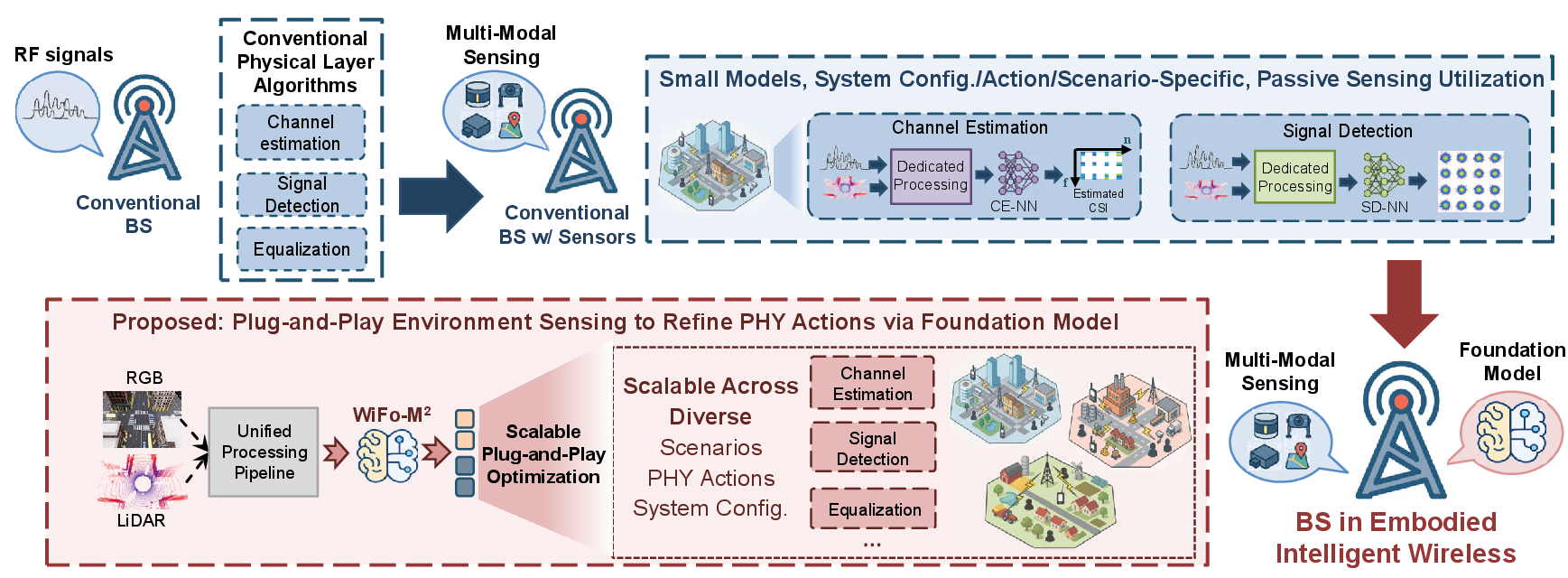}
    \caption{Comparisons between conventional RF-based/sensing-refined wireless communications paradigm, and the proposed FM-based approach of refining PHY actions with environment sensing under EIW paradigm.}
    \label{fig:enter-label}
    \vspace{-0.5em}
\end{figure*}

% However, the field of sensing-refined PHY actions is currently dominated by traditional per-task ``small models'' that achieve isolated improvements but lack a cohesive mechanism for intelligent entities to convert rich observations into consistently better PHY actions across diverse environments. As shown in Fig. \ref{fig:enter-label}, existing solutions are inherently PHY action-specific, system configuration-customized, and scenario-bound. 
% % However, two fundamental challenges hinder the wider adoption and effectiveness of multi-modal sensing in communication systems. 
% % First, the prevailing design paradigm remains largely task-specific and scenario-bound. 
% % , Existing approaches
% Specifically, they typically rely on meticulously customized data processing pipelines and network architectures tailored for narrow objectives, resulting in performance gains that lack universality and entail significant engineering overhead.   Moreover, current research has concentrated on a limited set of PHY actions (e.g., beam prediction and handover) with direct visual correlations, leaving broader actions (e.g., channel interpolation/prediction) untapped due to the more latent relationships between them and sensing.
% % leaving The potential of leveraging sensing to enhance a wider range of core transceiver modules, such as channel interpolation and prediction, remains largely untapped due to the complex and often latent relationships between sensing data and various transceiver modules.
However, as illustrated in Fig. \ref{fig:enter-label}, the field of sensing‑refined PHY actions remains dominated by traditional per‑task ``small models'' that deliver isolated gains but generalize poorly. They are scenario‑bound and configuration‑locked, lacking a cohesive mechanism to convert rich observations into consistently better PHY actions across environments. Moreover, the per-task paradigm keeps the research scope narrow and non-transferable. Each PHY action must be served by a customized data pipeline and model, driving up engineering cost and stalling large-scale deployment. Consequently, existing studies mostly address a few vision-friendly tasks such as beam prediction and handover, whereas broader needs like channel interpolation or prediction receive little attention. Finally, current BSs, though equipped with multi-modal sensors, still lack agentic‐AI capabilities. They passively fit sensing inputs to predefined outputs rather than learning latent scene dynamics and generally useful channel-aware features, leaving them unable to exploit observations under occlusion or to anticipate fine-grained  out-of-band (OOB) features.

To propel current sensing-refined PHY research toward the EIW paradigm and the vision of Agentic AI, we advocate a foundation model (FM) based approach that equips intelligent entities with a universal, task-agnostic observation-to-action bridge.
% Inspired by foundation models (FMs) \cite{FM,FM1},
In this paper, we introduce WiFo-M$^{2}$, the first universal FM that transforms multi-modal sensing into OOB channel-aware features, thereby serving as a plug-and-play performance booster across diverse PHY actions, as shown in Fig. \ref{fig:enter-label}. 
To establish positive sample pairs across heterogeneous modalities, we first develop a unified multi-modal sensing processing pipeline. {\color{black}We then design the ContraSoM pre-training strategy to align high-dimensional representations extracted from multi-modal sensing with the corresponding CSI features in a shared latent space. Moving beyond conventional static contrastive learning, ContraSoM introduces a continuous temporal extrapolation mechanism. By operating as a sequence-to-sequence model, WiFo-M$^2$ ingests historical sensory windows to extrapolate future channel-aligned features, effectively bridging the critical sampling rate gap between sparse sensing and dense PHY configuration frequency. Furthermore, to extract robust representations under data constraints, ContraSoM incorporates modality-specific augmentations, notably a diffusion-driven feature-space generation for Light Detection and Ranging (LiDAR) point clouds, which forces the network to explicitly learn channel-aware common structures.}
Our key contributions are summarized as follows:
\begin{itemize}
\item[ $\bullet$]
    We develop WiFo-M$^2$\footnote{Simulation codes are provided to reproduce the results presented in this paper: https://github.com/Haotian-Zhang-PKU/WiFoM2}, {\color{black}the first universal foundation model that pioneers a new paradigm for sensing-refined PHY actions. Moving beyond fragmented, task-specific networks, WiFo-M$^2$ establishes a unified OOB channel-aware feature space, empowering diverse downstream PHY actions in a highly scalable, plug-and-play manner without requiring manual architectural redesigns.}
\end{itemize}

\begin{itemize}
    \item[ $\bullet$]
    We propose a tailored pre-training strategy based on contrastive learning, named \textit{ContraSoM}. {\color{black}By co-designing a continuous temporal extrapolation mechanism and modality-specific data augmentations, ContraSoM endows WiFo-M$^2$ with the ability to bridge the critical inter-frame perception gap and extract robust, channel-aware common structures under limited paired data.}
 
\end{itemize}

\begin{itemize}
    \item[ $\bullet$]
    We validate the capabilities of WiFo-M$^2$ across extensive PHY actions implemented via diverse schemes. Experiments show that WiFo-M$^2$ empowers these PHY actions without requiring backbone fine-tuning and exhibits strong cross-scenario generalization capabilities.
\end{itemize}

\vspace{-0.6em}
\section{Related Work}
\subsection{Sensing-Refined PHY Actions}
Multi-modal sensing has been increasingly utilized to improve PHY actions. Existing schemes are predominantly action-specific, requiring carefully customized data processing and network designs for each application. For example, the LE-CLN scheme \cite{CLN} converts raw LiDAR data into the signal propagation feature superposed range map and customizes a feature extraction neural network (NN) to transform LiDAR into OOB features beneficial for channel estimation. Similarly, the H-MVMM scheme proposed in \cite{ICC} is specifically designed for precoding, with carefully crafted data processing methods for Global Positioning System (GPS), LiDAR, and RGB images, as well as customized feature extraction and fusion NNs. The MSCP channel prediction framework proposed in \cite{TCCN_CP} adopts a bespoke pipeline that first aligns camera images with pilots via angle transformation and path filtering, and then employs a task-specific cross-modal Transformer to regress channel parameters.
% The radar-based channel estimation method proposed in \cite{sensingaidedCE2022} extracts propagation delay, velocity, and angle parameters from radar data via clutter removal, 3D-FFT, and peak detection to guide sparse channel estimation.
Such design principle leads to a fundamental limitation: the performance gain lacks universality and is only effective for a particular PHY action, scenario, and transceiver setup.

Another pressing issue is that current research remains concentrated on just a few PHY actions whose outputs are strongly tied to user location, such as beam prediction and handover. To date, whether environment sensing can benefit many other PHY actions is still largely unexplored. The difficulty stems from the highly diverse and often latent relationships between sensing and PHY actions. For example, temporal channel prediction (CP) relies on capturing time-varying characteristics of scatterers; uplink–downlink CP concerns spatial consistency across frequency bands. Hence, crafting a dedicated sensing pipeline and mining task-specific features becomes labor-intensive.
% In this paper, we demonstrate for the first time, through the proposed WiFo-M$^2$, that multi-modal sensing can deliver significant gains for both channel prediction and channel interpolation tasks. This broadens the application scope of multi-modal sensing, extending its benefits beyond the traditionally limited set of optimized functions to encompass a wider range of core transceiver modules. 
In this context, how to obtain a universal OOB feature from multi-modal sensing and use it to boost a wide range of PHY actions becomes a critical challenge.
% This study is the first to demonstrate that multi-modal sensing can boost channel interpolation and channel prediction via WiFo-M$^2$, two core transceiver modules that had not been explored for such benefits until now.

% \vspace{-1.0em}
\subsection{CL-Based Foundation Model for Cross-Modal Alignment}
WiFo-M$^2$ is envisioned to extract universal OOB channel-aware features from multi-modal environment sensing. Achieving this requires a self-supervised pre-training scheme that (i) learns from large-scale unlabeled data and (ii) aligns heterogeneous modalities.
Contrastive learning (CL)~\cite{CLIP}, which constructs representations by learning similarities between data samples, naturally meets the above requirements. Recently, beyond traditional vision-language domains, CL has been effectively adapted to align heterogeneous representations across diverse and complex application scenarios.
% UNIMO \cite{ACL} introduces a unified architecture that leverages cross-modal CL to align visual and textual representations not only with paired data but also by incorporating augmented single-modal data, enhancing the generality of learned features for both understanding and generation tasks. In multimodal recommendation, 
DiffCL~\cite{tnnls} introduces a diffusion-based graph CL module that generates denoised visual- and text-aware views, encouraging the encoder to distill robust user–item representations even under severe interaction sparsity and multi-modal noise.
% In 3D object detection, CAT-Det \cite{CAT_Det} employs hierarchical contrastive learning to align LiDAR point clouds with RGB images, using modality-specific augmentations to overcome data misalignment and improve detection robustness.
In medical image segmentation, CLMorph~\cite{tnnls1} integrates CL into a deformable registration framework to align an unaligned image with a reference image. By contrasting feature representations extracted from these images, it enhances the discriminative capacity of the network. 
% In industrial time-series generation, MetaIndux-TS~\cite{tnnls2} employs a contrastive synthesis layer within its diffusion framework to enhance generation fidelity. This layer explicitly compares the synthesized sequences with the original noisy input, guiding the model to better capture fine-grained deviations and improve the overall quality of generated data. 
Similarly, in multi-media recommendation, graph-based contrastive methods \cite{SIGIR} are designed to disentangle and align user preferences across visual, acoustic, and textual modalities.

Building upon these advances, CL has also been introduced as an effective pre-training strategy in wireless communication community. CSI-CLIP proposed in \cite{CL_CSICIR} treats CSI and channel impulse response as naturally aligned pairs and performs cross-modal CL between them, enabling the model to capture generalizable channel features for improved positioning and beam management. LWLM~\cite{CL_LWLM} introduces position-invariant CL, which contrasts CSI samples collected from the same location under different BS configurations to learn robust and location-aware semantics specifically for localization tasks. Furthermore, in~\cite{CL_beam}, a VLM-based beam prediction (BP) framework is proposed which utilizes CL to align image and LiDAR features, enforcing cross-modal consistency to enhance multi-modal fusion for beam selection. 
% ContraWiMAE~\cite{CL_wireless} proposes a hybrid framework that combines CL with masked reconstruction. It constructs positive pairs by injecting noises into channel samples, enabling the model to jointly learn discriminative and structural features.
Most recently, the WMFM model \cite{WMFM} adopts a vanilla form of CL to align images with CSI, yet it still suffers from several limitations. Firstly, as in most prior work, it fails to resolve the intrinsic temporal mismatch between sensory data and CSI. Secondly, it utilizes solely the image modality and does not employ any data pre-processing and augmentation techniques, limiting its robustness and cross-domain generalization. Thirdly, the pre-training dataset covers only one scene and a single communication system configuration. Finally, it does not investigate how multi-modal sensing could benefit a wider range of PHY actions and remains confined to Line of sight (LoS)/non-Line of sight (NLoS) classification and localization.

\vspace{-0.5em}
\section{System Model}
\begin{figure}
    \centering
    \includegraphics[width=0.98\linewidth]{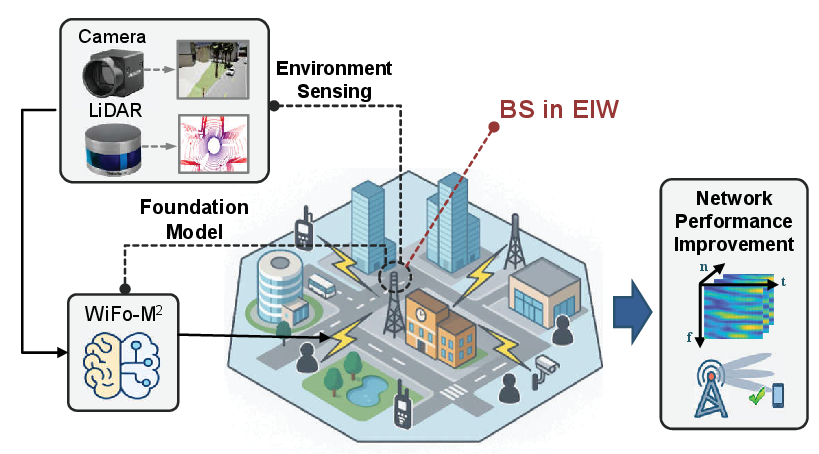}
    \caption{An overview of the considered scenario.}
    \label{fig:sm}
    \vspace{-0.6em}
\end{figure}

% \subsection{System Model}
As illustrated in Fig.~\ref{fig:sm}, we consider an intelligent transportation scenario, where the BS is an intelligent agent that serves $M$ single‑antenna users and senses the physical world through multi‑modal sensors. In this work, we assume that multi‑modal environment sensing includes LiDAR point clouds and RGB images, denoted by $\bm{P}_{\text{raw}} \in \mathbb{R}^{N \times 3}$ and $\bm{I} \in \mathbb{R}^{3 \times H \times W}$, respectively. We assume the camera operates at $40$ fps, while the LiDAR scans at $20$ Hz.

We consider a multiple-input single-output orthogonal frequency division multiplexing (MISO-OFDM) communication system\footnote{{\color{black}Because WiFo-M$^2$ extracts universal channel-aware features derived directly from the physical space, it is fundamentally decoupled from specific antenna arrays and can seamlessly adapt to multiple input multiple output (MIMO), single input multiple output (SIMO), and single input single output (SISO) systems.}} under EIW paradigm where BS is equipped with either a uniform linear array (ULA) or a uniform planar array (UPA), with the antenna size denoted by $N$ and the number of subcarriers by $K$. Let $\mathbf{H} \in\mathbb{C}^{K \times N}$  represent CSI between the BS and a certain user.

\begin{figure*}
    \centering
    \includegraphics[width=1\linewidth]{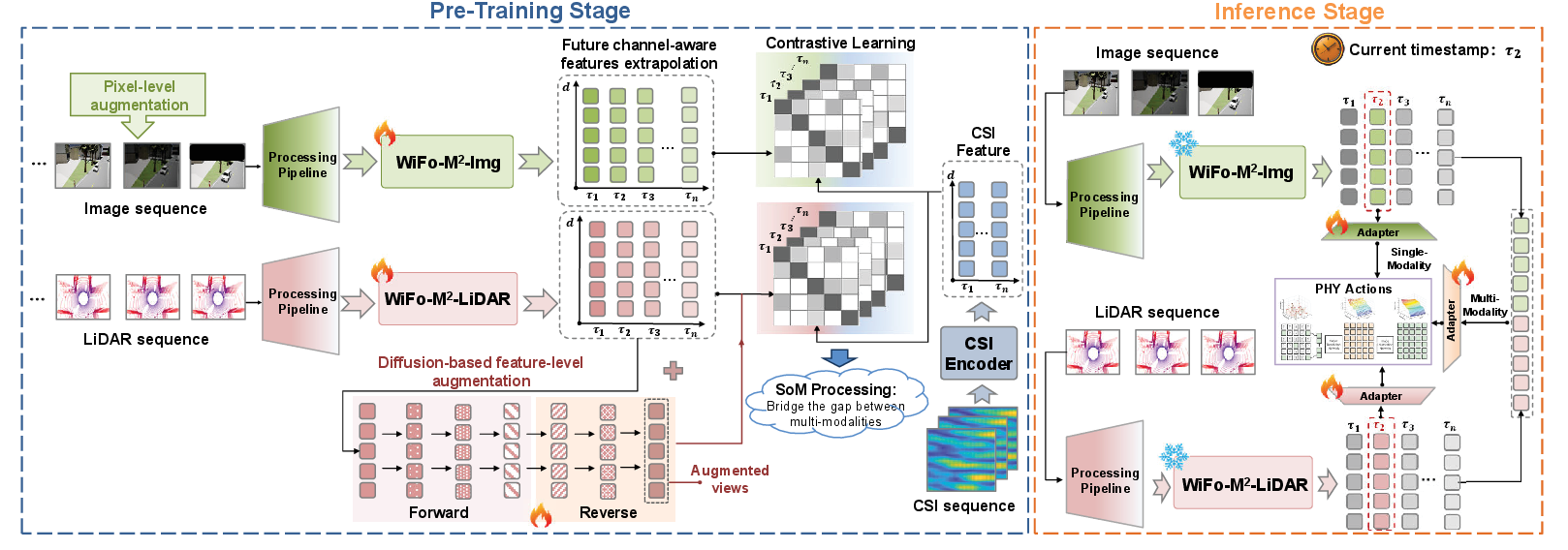}
    \caption{Overall framework of WiFo-M$^2$, illustrating the two‑stage pipeline: (left) pre-training stage with ContraSoM strategy realizing SoM processing, and (right) inference stage for optimizing various PHY actions in a plug-and-play manner.}
    \label{fig:detail}
    % \vspace{-0.2em}
\end{figure*}

\section{WiFo-M$^2$ Framework: Sensing Processing, Feature Extraction, and Pre-training}
In this section, we first analyze the fundamental challenges faced by developing WiFo-M$^2$ to achieve scalable environment sensing-refined PHY actions for intelligent network entities. Then, we introduce the core designs: the unified environment sensing processing pipeline, the architecture and feature extraction method of the WiFo-M$^2$, and the ContraSoM pre‑training strategy. The overall framework of WiFo‑M$^2$ is depicted in Fig. \ref{fig:detail}.
% \vspace{-0.8em}
\subsection{Challenges}
% Substantial modifications are needed in both the pre-training strategy and the model architecture to cope with the unique challenges posed by multi-modal sensing and wireless channels. The key obstacles are:
To realize the goal of scalable sensing-refined PHY actions, the primary obstacles arise from three key aspects: cross-modal data pairing, temporal-aware model design, and robust pre-training strategies. These are detailed as follows:

\textit{(1) Absence of Inherent Cross-Modal Data Pairing:} 
A fundamental premise for CL is the availability of aligned positive sample pairs across modalities. 
In wireless communication scenarios, however, a single sensing frame captures elements like vehicles and buildings can correspond to multiple potential receivers. Consequently, the first challenge lies in disentangling and constructing matching CSI positive pairs for different receivers from the same set of sensory data. 

\textit{(2) Temporal Misalignment Due to Heterogeneous Sampling Rates:} 
% The goal of sensing-aided communications is to provide OOB channel feature at the same measurement rate as the CSI estimation itself. However, a fundamental gap exists: 
Wireless channels vary every few milliseconds, whereas intelligent network entities can only observe the surrounding environment at low frame rates (e.g., $10$-$20$ Hz). {\color{black}Notably, this critical physical reality has been largely ignored by prevailing sensing-aided PHY research, which often unrealistically assumes perfect temporal alignment \cite{TCCN_CP,TCCN_CE,CLN,MMFF,ICC}.}
Therefore, most PHY actions lack concurrent environmental observations during execution.
Therefore, WiFo-M$^2$ cannot simply rely on aligning sensory frames with coincident CSI samples.  Instead, it has to effectively ``fill in” the missing environmental context at the communication system's operational tempo.

{\color{black}\textit{(3) Extracting Modality-Invariant Features under Data Limitation:} Data augmentation is crucial for generating diverse, semantic-preserving views to learn robust representations. However, crafting augmentation strategies for complex, unstructured modalities like LiDAR point clouds is non-trivial. Beyond simply expanding the training volume, the fundamental challenge is to design controlled perturbations that force the network to discard environmental details irrelevant to signal propagation and explicitly learn the underlying channel-aware common structures, ensuring the generated views remain valid positive samples for contrastive learning.}

{\color{black}While multi-modal CL has achieved remarkable success, directly adopting vanilla CL paradigms is fundamentally insufficient to tackle the aforementioned challenges in wireless communications. Standard CL inherently assumes naturally synchronized and densely paired static multi-modal inputs. However, practical communication systems suffer from severe modality sampling rate discrepancies and extreme data collection costs. To bridge these gaps, our proposed ContraSoM framework significantly customizes the conventional CL philosophy. Specifically, rather than relying on instantaneous static alignment, WiFo-M$^2$ incorporates continuous temporal extrapolation to perform global temporal CL against sequential CSI states. Furthermore, to combat the inherent scarcity of sensing-communication datasets, we integrate modality-customized feature augmentations, leveraging pixel transformations for visual streams and diffusion mechanisms for LiDAR point clouds. Consequently, these tailored innovations empower ContraSoM to extract robust, sampling-rate-agnostic, and highly generalizable OOB channel-aware representations. }

% \vspace{-2.6em}
\subsection{Unified Environment Sensing Processing Pipeline}
\subsubsection{Image Processing Pipeline}
The image processing pipeline converts raw images containing many potential receivers into a representation deterministically linked to the CSI of a specific transmitter and receiver pair, and it comprises three core steps.

\textbf{Step 1: Initial Object Detection and Tracking.} The raw image data is first subjected to a pixel-level augmentation pipeline (described in detail in Section \ref{MSDA}), and the resulting image is denoted as $\tilde{\bm{I}}$. The image processing pipeline begins with object detection to identify potential communication targets (e.g., vehicles in V2X communication scenarios) from $\tilde{\bm{I}}$. We assume the BS has prior knowledge of the receiver's initial position through initial access procedures. Then, the advanced object detection model YOLOv8 is used to detect all targets in the image and perform continuous tracking across frames.

\textbf{Step 2: Receiver-Bounding Box Matching.} For each detected bounding box, we calculate its angular span in the camera's field of view (FoV).  Given a bounding box with normalized coordinates $(x_c^n, y_c^n, w^n, h^n)$ where $x_c^n, y_c^n$ are the normalized center coordinates and $w^n, h^n$ are the normalized width and height, we first convert to pixel coordinates: $x_c = x_c^n \times W$, $y_c = y_c^n \times H$, $w = w^n \times W$, and $h = h^n \times H$, where $W$ and $H$ are the image width and height in pixels. The horizontal angular span $[\theta_{\text{min}}, \theta_{\text{max}}]$ of the bounding box is then computed using the camera's intrinsic parameters:
{
\setlength{\abovedisplayskip}{3pt}%
  \setlength{\belowdisplayskip}{3pt}%
\begin{align}
\theta_{\text{min}} &= \arctan\left(\frac{x_{\text{min}} - c_x}{f_x}\right), \\
\theta_{\text{max}} &= \arctan\left(\frac{x_{\text{max}} - c_x}{f_x}\right),
\end{align}
}where $x_{\text{min}} = x_c - w/2$, $x_{\text{max}} = x_c + w/2$, $f_x$ is the focal length in pixels, and $c_x$ is principal point offset along x-axis.

To match the receiver with the correct bounding box, we compute the receiver's azimuth angle relative to camera. Given the camera's world coordinates $\bm{p}_{\text{cam}} = (x_{\text{cam}}, y_{\text{cam}}, z_{\text{cam}})$ and yaw angle $\psi_{\text{cam}}$, and the receiver's world coordinates $\bm{p}_{\text{rec}} = (x_{\text{rec}}, y_{\text{rec}}, z_{\text{rec}})$, the relative azimuth angle is:
\begin{equation}
\theta_{\text{rec}} = \text{wrap}\left(\arctan\left(\frac{y_{\text{rec}} - y_{\text{cam}}}{x_{\text{rec}} - x_{\text{cam}}}\right) - \psi_{\text{cam}}\right),
\end{equation}
where $\text{wrap}(\cdot)$ ensures the angle remains within $(-\pi, \pi]$. The bounding box whose angular span contains $\theta_{\text{rec}}$ is selected as the match. If multiple bounding boxes satisfy this condition, we choose the one whose angular center $\bar{\theta} = (\theta_{\text{min}} + \theta_{\text{max}})/2$ is closest to $\theta_{\text{rec}}$.

\textbf{Step 3: Receiver-Labeling via Angular Encoding.} Once the matching bounding box is identified, we perform a receiver-labeling operation by replacing the original pixel content within the bounding box with a color that encodes the receiver's angular information. Through this operation, a positive sample association between the processed image and the CSI between BS and a specific user is established, and geometric information is also directly embedded into the image for subsequent feature learning.

The angular encoding proceeds as follows. First, we clip the receiver's azimuth angle to the range $[-\pi/2, \pi/2]$, i.e., $\theta_{\text{clipped}} = \max\left(-\frac{\pi}{2}, \min\left(\frac{\pi}{2}, \theta_{\text{rec}}\right)\right)$. Next, we normalize this angle to the range $[0, 1]$ by $\theta_{\text{norm}} = (\theta_{\text{clipped}} + \pi/2) / \pi$. Finally, we map the normalized angle to a hue value in the HSV color space via $h = \lfloor 179 \times \theta_{\text{norm}} \rfloor$. The corresponding RGB color $\bm{c}_{\text{RGB}} = (r, g, b)$ is obtained by converting the HSV color $(h, 255, 255)$ to the RGB color space. The bounding box region is then filled with this color: 
{
\setlength{\abovedisplayskip}{3pt}%
  \setlength{\belowdisplayskip}{3pt}%
\begin{equation}
    \tilde{\bm{I}}(y_1:y_2, x_1:x_2) = \bm{c}_{\text{RGB}},
\end{equation}}where $(x_1, y_1)$ and $(x_2, y_2)$ define the bounding box corners in pixel coordinates. The transformed image after this operation is denoted as $\bm{I}_{\rm labeled}$.

% This processing pipeline transforms raw images into **receiver-labeled images** that retain geometric relationships while abstracting away unnecessary visual details, creating an effective input representation for subsequent feature extraction and multimodal alignment.
\subsubsection{LiDAR Processing Pipeline}
Given a raw point cloud $\bm{P}_{\text{raw}} \in \mathbb{R}^{N \times 3}$ consisting of $N$ points with 3D coordinates $(x, y, z)$, we first remove ground points and obtain a filtered point cloud $\bm{P}_{\text{filt}} \in \mathbb{R}^{N' \times 3}$.

\textbf{Step 1: Object Clustering and Initial Receiver Matching.} To segment the filtered point cloud into distinct objects, we apply the DBSCAN clustering algorithm to $\bm{P}_{\text{filt}}$, obtaining a set of clusters $\mathcal{C} = \{C_1, C_2, \dots, C_K\}$. For each cluster $C_k$, we compute its centroid $\bm{c}_k = (x_k, y_k, z_k)$. During the initial access phase, the receiver reports its global position $\bm{p}_{\text{rec}} = (x_{\text{rec}}, y_{\text{rec}}, z_{\text{rec}})$. The receiver‑corresponding cluster is identified as the one whose centroid is closest to $\bm{p}_{\text{rec}}$:
$k^* = \arg\min_{k} \|\bm{c}_k - \bm{p}_{\text{rec}}\|$.

\textbf{Step 2: Continuous Tracking and Point Labeling.} Once the receiver cluster is identified in the initial frame, we employ a multi‑object tracking algorithm to maintain consistent cluster identities across consecutive frames. This ensures that the receiver vehicle is correctly associated throughout the sequence. Let the receiver cluster be denoted as $\mathcal{C}_{\text{rec}}$, and let the set of building points be defined by a geometric rule, e.g., \(\mathcal{B} = \{\bm{p}_i \in \bm{P}_{\text{filt}} : y_i > y_{\text{build}}\}\). Then each point \(\bm{p}_i = (x_i, y_i, z_i)\) in \(\bm{P}_{\text{filt}}\) is assigned a label \(l_i \in \{-1, 0, 1\}\) according to:
{
\begin{align}
l_i = 
\begin{cases}
1, & \text{if } \bm{p}_i \in \mathcal{C}_{\text{rec}} \\
-1, & \text{if } \bm{p}_i \in \mathcal{B} \setminus \mathcal{C}_{\text{rec}} \\
0, & \text{otherwise}
\end{cases}.
\end{align}}

Finally, the labeled point cloud \(\bm{P}_{\text{labeled}} \in \mathbb{R}^{N' \times 4}\) is then constructed by appending the label column to \(\bm{P}_{\text{filt}}\), i.e., \(\bm{P}_{\text{labeled}} = [\bm{P}_{\text{filt}}, \bm{l}]\), where \(\bm{l} = [l_1, l_2, \dots, l_{N'}]^T\).

{\color{black}
\textit{Remark}: It is worth noting that in practical scenarios, object detection or tracking may occasionally fail due to severe occlusion or algorithmic limitations. In such cases, the sensory data is processed without receiver-labeling. Crucially, these unlabeled frames are still retained and paired with the corresponding CSI as positive samples during the ContraSoM pre-training. Given that the wireless channel is governed by both dynamic scatterers and static backgrounds (e.g., buildings and roads), the unlabeled sensory data still provides valuable information by capturing the static propagation environment~\cite{Ref_Xing}. Consequently, contrastive learning on these unlabeled pairs explicitly forces WiFo-M$^2$ to learn scenario-general priors and static channel components.
}

\subsection{Network Architecture and Working Flow of WiFo-M$^2$}
\begin{figure}
    \centering
    \includegraphics[width=0.96\linewidth]{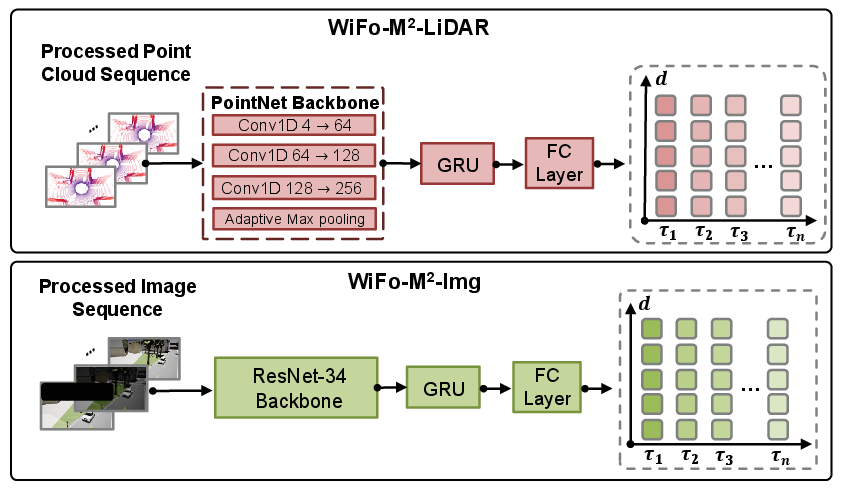}
    \caption{Network architecture of WiFo-M$^2$-Img/LiDAR.}
    \label{fig:net}
    \vspace{-0.6em}
\end{figure}

{\color{black}
WiFo-M$^2$ is a collective name representing a suite of modality-specific foundation models. In this paper, it encompasses WiFo-M$^2$-Img (built upon a ResNet-34 backbone \cite{resnet}) and WiFo-M$^2$-LiDAR (built upon a PointNet backbone \cite{pointnet}). Their architectures are depicted in Fig.~\ref{fig:net}. While this work instantiates WiFo-M$^2$-Img and WiFo-M$^2$-LiDAR for evaluation, the framework is inherently flexible and capable of encompassing other modality-specific sub-models to accommodate a broader range of sensors. Unlike traditional networks designed for specific tasks, the core function of these WiFo-M$^2$ models is to serve as channel feature learners. Their role is to distill universal, OOB channel-aware features from raw multi-modal sensory data, which are subsequently utilized to empower diverse downstream PHY actions.}
\subsubsection{CSI Feature Extraction}
For CSI feature extraction, we employ the pre-trained WiFo model \cite{wifo} as a powerful space-temporal-frequency (STF) feature extractor. Given an input CSI data $\mathbf{H}_t \in \mathbb{C}^{K \times N}$ at timestamp $t$, the WiFo model processes it through its transformer-based encoder. 
% During feature extraction, we set the masking ratio to zero, allowing the model to function as a comprehensive CSI encoder without reconstruction objectives.
The output is a fixed-dimensional feature vector $\mathbf{z}^{\text{C}}_t \in \mathbb{R}^{d}$.
\subsubsection{Image and LiDAR Feature Extraction}
Both image and LiDAR modalities follow the same feature extraction procedure comprising three stages:

\textbf{Stage 1: Spatial Feature Extraction.} Let $T$ denote the timestamp of the most recent sensory data. Define $\Delta t_{\rm I}$ and $\Delta t_{\rm L}$ as the inter-frame intervals of the image and LiDAR streams, respectively, and $(n_{\rm I}+1)$ and $(n_{\rm L}+1)$ as the sequence lengths of the LiDAR and image inputs. For image modality, each frame $\bm{I}^t_{\rm labeled}$ passes through the ResNet-34 backbone of the WiFo-M$^2$-Img model to obtain frame-level features. For LiDAR modality, each point cloud frame $\bm{P}^t_{\rm labeled}$ is processed by the PointNet backbone \cite{pointnet} to extract frame-level features. The extraction process can be formally expressed as:
\begin{align}
\bm{u}_t^{\text{I}} &= f_{\text{ResNet}}\left(\bm{I}^t_{\rm labeled}\right) \in \mathbb{R}^{d}, \  t = T-n_{\rm I}\Delta t_{\rm I}, \ldots, T \\
\bm{u}_t^{\text{L}} &= f_{\text{PointNet}}\left(\bm{P}^t_{\rm labeled}\right) \in \mathbb{R}^{d}, \  t = T-n_{\rm L}\Delta t_{\rm L}, \ldots, T
\end{align}
where $f_{\text{ResNet}}$ and $f_{\text{PointNet}}$ denote the ResNet-34 and PointNet backbone networks respectively. We denote the frame-level features for image and LiDAR within a time window $[T-n_{\rm I}\Delta t_{\rm I},T]$ or $[T-n_{\rm L}\Delta t_{\rm L},T]$ as $\bm{u}_t^{\text{I}} \in \mathbb{R}^{d}$ and $\bm{u}_t^{\text{L}} \in \mathbb{R}^{d}$, respectively.

\textbf{Stage 2: Temporal Feature Modeling.} To capture temporal dependencies, both modalities employ the GRU network that learns the underlying dynamics of the environment. The GRU processes the sequences of intermediate frame-level features $\{\bm{u}^{\rm x}_t\}_{t=T-n_{\rm x}\Delta t_{\rm x}}^{T}, {\rm x}\in\{\rm L, \rm I\}$ to produce a sequence of hidden states, which are then projected to the final feature space. Let $\Delta \tau$ be the PHY action execution interval. The resulting projected feature sequence corresponds to the prediction window $[T,T+\Delta \tau,\cdots,T+\Delta t_{\rm x}]$, i.e., from the current sensing timestamp $T$ up to the moment when the next sensing frame is captured. This operation can be expressed as:
\begin{align}
\bm{v}^{\rm x}_\tau\ &= \phi_{\rm x}\Bigl(\mathcal{G}_{\rm x}\left(\{\bm{u}^{\rm x}_t\}_{t=T-n_{\rm x}\Delta t_{\rm x}}^{T}\right)\Bigr), \tau=T, \cdots,T+\Delta t_{\rm x} 
% \\
% \bm{v}^{\rm L}_\tau\ &= \phi_{\rm L}\Bigl(\mathcal{G}_{\rm L}\left(\{\bm{u}^{\rm L}_t\}_{t=T-n_{\rm L}\Delta t_{\rm L}}^{T}\right)\Bigr),  \tau=T, \cdots,T+\Delta t_{\rm L} 
\end{align}
where $\mathcal{G}_{\rm x}$ denotes the GRU network for modality $\rm x$, $\phi_{\rm x}$ represents the projection layer, and $\bm{v}_t^{\rm x} \in \mathbb{R}^{d}$ is the final frame-level feature at time step $t$ for modality $\rm x$. The sequence of final features $\{\bm{v}^{\rm x}_\tau\}_{\tau=T}^{T+\Delta t_{\rm x}}$ is then used for subsequent temporal alignment and contrastive learning.

% These temporal models enable the model to learn the underlying dynamics of the environment and extrapolate features to future time steps.
\begin{figure}
    \centering
    \includegraphics[width=1\linewidth]{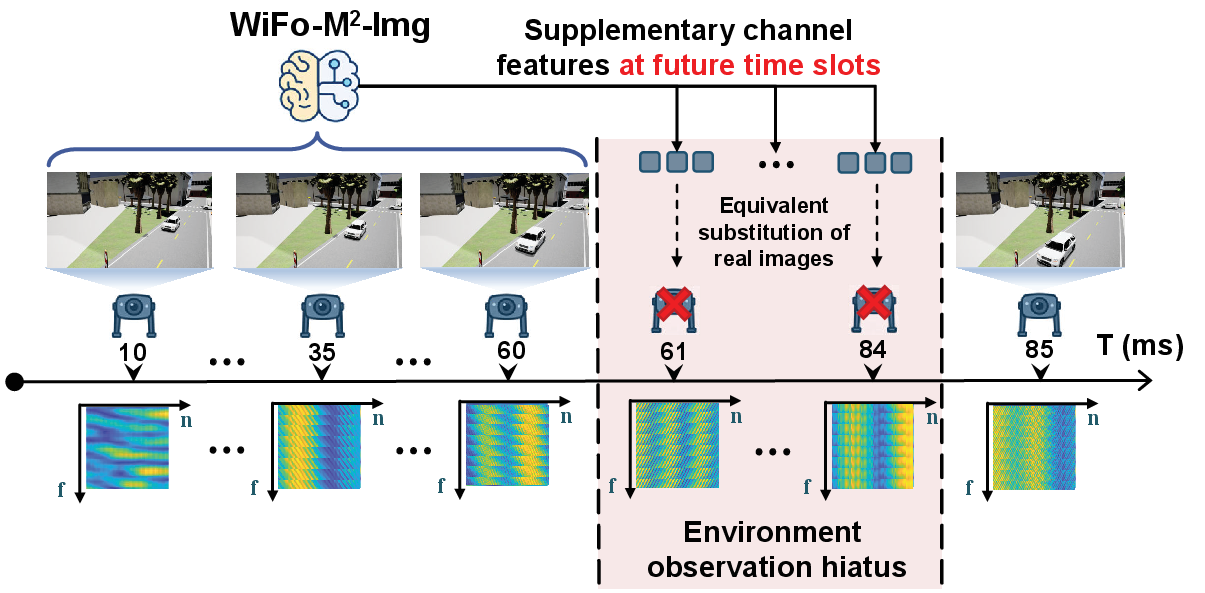}
    \caption{{\color{black}Temporal feature extrapolation in WiFo-M$^2$-Img.}}
    \label{fig:time_sequence}
    % \vspace{-0.3em}
\end{figure}
\textbf{Stage 3: Multi-Timestamp Feature Generation and Alignment.} A key design of WiFo-M$^2$'s architecture is that the GRU generates channel-aware features for multiple future timestamps. During pre-training, the sequence of final features $\{\bm{v}^{\rm x}_\tau\}_{\tau=T}^{T+\Delta t_{\rm x}},{\rm x}\in\{\rm L, \rm I\}$ generated by WiFo-M$^2$ are individually aligned with the CSI feature at the corresponding timestamp. For a given timestamp $\tau$, a certain sensing feature $\bm{v}^{\rm{x}}_{\tau}$ is selected and paired with the CSI feature $\bm{z}^{\text{C}}_{\tau}$ from the same timestamp to form a positive pair for CL (detailed in the next section). \textcolor{black}{Crucially, this mechanism ensures that the GRU extrapolation does not operate as an unconstrained approximation. Rather, each extrapolated feature is strictly supervised by actual fine-grained CSI. As shown in Fig.~\ref{fig:time_sequence}, this strict supervision explicitly forces WiFo-M$^2$ to capture the true dynamic evolution of the channel, ensuring physically valid and precise cross-modal temporal alignment.}

\vspace{-0.5em}
\subsection{Pre-training Strategy for WiFo-M$^2$: ContraSoM}
This section presents the proposed ContraSoM pre-training strategy in detail, covering the modality-specific data-augmentation techniques together with the overall CL pipeline. By aligning multi-modal environment sensing and wireless channel in the latent space, ContraSoM essentially performs the SoM processing outlined in \cite{SoM}.
{\color{black}Note that the proposed ContraSoM pre-training strategy is inherently modality-agnostic, ensuring the high extensibility of the WiFo-M$^2$ framework. To incorporate additional sensing modalities such as Radar or GPS, one simply needs to integrate a corresponding modality-specific backbone for initial feature extraction, and subsequently align these features with the CSI using the same pipeline.}

\subsubsection{Modality-Specific Data Augmentation}
\label{MSDA}
Image data possess a regular grid structure and dense pixel representation, making them amenable to direct augmentation in the pixel space. We apply a series of stochastic transformations to generate robust augmented views for contrastive learning. Given an input image $\bm{I}$, the augmented view $\tilde{\bm{I}}$ is obtained via: $\tilde{\bm{I}} = \mathcal{T}_{\text{aug}}(\bm{I})$.  $\mathcal{T}_{\text{aug}}$ is a combination of several of the following operations, randomly applied to each original image: 
% Firstly, the image is resized to $224 \times 224$ using bilinear interpolation.
First, color jitter randomly perturbs brightness, contrast, saturation, and hue, mimicking natural variations in illumination: $\bm{I}_{\text{jit}} = \bm{I} \odot \mathbf{c}_{\text{scale}} + \mathbf{c}_{\text{shift}}$, where $\odot$ denotes element-wise multiplication, $\mathbf{c}_{\text{scale}} \sim \mathcal{U}(0.7,1.3)$ and $\mathbf{c}_{\text{shift}} \sim \mathcal{U}(-0.3,0.3)$ for each channel. Second, Gaussian Blur is applied with probability $p=0.5$ to simulate lens defocus or motion blur: $\bm{I}_{\text{blur}} = \bm{I} * G_{\sigma}$ with $\sigma \sim \mathcal{U}(0.1, 2.0)$, where $*$ denotes the convolution operation and $G_{\sigma}$ is a Gaussian kernel of size $15 \times 15$. Next, random erasing is performed with probability $p=0.3$, which randomly selects a rectangular region $\mathcal{R}$ within the image and replaces the pixel values in that region with random noise, emulating occlusions or sensor dropout. This operation is defined as: 
{
\begin{equation}
    \bm{I}_{\text{erase}}(x,y) = \begin{cases}
\eta, & \text{if } (x,y) \in \mathcal{R} \\
\bm{I}(x,y), & \text{otherwise}
\end{cases},
\end{equation}}where $\eta \sim \mathcal{U}(0,1)$, and the region $\mathcal{R}$ has an area ratio $\sim \mathcal{U}(0.01,0.05)$ and an aspect ratio $\sim \mathcal{U}(0.3,3.3)$.  Finally, normalization is applied using ImageNet statistics: $\bm{I}_{\text{norm}} = (\bm{I} - \boldsymbol{\mu}) / \boldsymbol{\sigma}$ with $\boldsymbol{\mu} = [0.485, 0.456, 0.406]$ and $\boldsymbol{\sigma} = [0.229, 0.224, 0.225]$. This pipeline encourages the model to learn invariant representations under appearance variations, occlusion, and noise.

LiDAR point clouds are sparse and geometrically structured. Applying noise directly in the raw data level would easily disrupt semantic relationships. Therefore, we propose to perform augmentation in the feature space. We adopt a {denoising diffusion probabilistic model (DDPM)} framework to generate augmented features in a controlled manner. 

The augmentation process begins with a forward noising step. Taking the final LiDAR feature at a specific time step $i$, denoted as $\bm{v}^{\rm L}_{i}$, as an example, we gradually inject Gaussian noise into this feature over a series of $M$ discrete steps. This forward noising process is defined as a Markov chain:
{
\begin{equation}
    q(\bm{v}_m \mid \bm{v}^{\rm L}_{m-1}) = \mathcal{N}\left(\bm{v}^{\rm L}_m; \sqrt{1-\beta_m} \, \bm{v}^{\rm L}_{m-1}, \beta_m \mathbf{I} \right),
\end{equation}}where $\beta_m$ follows a linearly increasing schedule from $\beta_{\min} = 5\times10^{-6}$ to $\beta_{\max} = 2\times10^{-3}$, $m=1,\dots,M$. Due to the additive property of Gaussian noise, we can directly sample the feature at any step $m$:
\begin{equation}
\label{add_noise}
    \tilde{\bm{v}}^{\rm L}_m = \sqrt{\bar{\alpha}_m} \, \bm{v}^{\rm L}_{i} + \sqrt{1-\bar{\alpha}_m} \, \epsilon, \quad \epsilon \sim \mathcal{N}(0,\mathbf{I})
\end{equation}
with $\alpha_m = 1-\beta_m$ and $\bar{\alpha}_m = \prod_{s=1}^m \alpha_s$.

{For efficient feature generation in the reverse process, we use a deterministic skip-step reverse diffusion sampler. Starting from a noisy feature \(\tilde{\bm{v}}^{\rm L}_m\), we update with stride \(\Delta m = \max (\lfloor m / u \rfloor, 1)\) (where \(u=8\)) as
\begin{align}
\nonumber
\tilde{\bm{v}}^{\rm L}_{m-\Delta m} = &\sqrt{\bar{\alpha}_{m-\Delta m}} \left( \frac{\tilde{\bm{v}}^{\rm L}_m - \sqrt{1-\bar{\alpha}_m} \, \epsilon_\theta(\tilde{\bm{v}}^{\rm L}_m, m)}{\sqrt{\bar{\alpha}_m}} \right) \\
&+ \sqrt{1-\bar{\alpha}_{m-\Delta m}} \, \epsilon_\theta(\tilde{\bm{v}}^{\rm L}_m, m).
\end{align}
% where the noise predictor $\epsilon_\theta(\cdot)$ is a learnable NN.} Starting from a randomly sampled noisy feature $\tilde{\bm{v}}^{\rm L}_m$, the DDIM sampler with stride $\Delta m = \max (\lfloor m / u \rfloor,1)$ (where $u=8$) efficiently recovers a cleaned feature in fewer steps:
% \begin{align}
% \nonumber
% \tilde{\bm{v}}^{\rm L}_{m-\Delta m} = &\sqrt{\bar{\alpha}_{m-\Delta m}} \left( \frac{\tilde{\bm{v}}^{\rm L}_m - \sqrt{1-\bar{\alpha}_m} \, \epsilon_\theta(\tilde{\bm{v}}^{\rm L}_m, m)}{\sqrt{\bar{\alpha}_m}} \right) \\
% &+ \sqrt{1-\bar{\alpha}_{m-\Delta m}} \, \epsilon_\theta(\tilde{\bm{v}}^{\rm L}_m, m).
% \end{align}
After $U = 24$ reverse steps, we obtain the augmented feature $\tilde{\bm{v}}^{\rm L}_{i}$. To enhance diversity, we generate {two distinct augmented views} $\tilde{\bm{v}}_{i}^{\rm L(1)}$ and $\tilde{\bm{v}}_{i}^{\rm L(2)}$ by initializing from different noise levels and noise draws, then applying the same reverse update:
\begin{align}
\tilde{\bm{v}}^{\rm L}_{m_j} = \sqrt{\bar{\alpha}_{m_j}} \, \bm{v}^{\rm L}_{i} + \sqrt{1-\bar{\alpha}_{m_j}} \, \epsilon_j, \quad j\in\{1,2\},
\end{align}
where $m_1, m_2 \sim \mathcal{U}\{200, 400\}$ and $\epsilon_1, \epsilon_2 \sim \mathcal{N}(0,\mathbf{I})$. Then, run the above deterministic skip-step reverse updates for \(U\) steps to obtain
\[
\tilde{\bm{v}}_{i}^{\rm L(1)} = \mathcal{G}_{\text{rev}}(\tilde{\bm{v}}^{\rm L}_{m_1}), 
\qquad 
\tilde{\bm{v}}_{i}^{\rm L(2)} = \mathcal{G}_{\text{rev}}(\tilde{\bm{v}}^{\rm L}_{m_2}),
\]
where \(\mathcal{G}_{\text{rev}}(\cdot)\) denotes the \(U\)-step deterministic reverse sampler defined above. The two independent pairs \((m_1,\epsilon_1)\) and \((m_2,\epsilon_2)\) yield two complementary augmented views.

\subsubsection{Contrastive Learning Objectives in WiFo-M$^2$}
{\color{black}
To align the multi-modal environment sensing with CSI in the latent space, we design two distinct CL frameworks for {WiFo-M$^2$-Img} and WiFo-M$^2$-LiDAR. Before detailing the modality-specific objectives, we first introduce the preliminary formulation of the contrastive loss used in our framework.

The foundation of our contrastive alignment is the InfoNCE loss. For a batch of $N$ feature pairs from two distinct sets $Z_{a}$ and $Z_{b}$, the standard InfoNCE loss is defined as:
\begin{equation}
    \mathcal{L}_{\text{InfoN}}(\mathbf{Z}_a, \mathbf{Z}_b) = -\frac{1}{N} \sum_{i=1}^{N} \log \frac{\exp\left(s(\mathbf{z}_a^{(i)}, \mathbf{z}_b^{(i)}) / \tau_{\rm e}\right)}{\sum_{j=1}^{N} \exp\left(s(\mathbf{z}_a^{(i)}, \mathbf{z}_b^{(j)}) / \tau_{\rm e}\right)},
\end{equation}
where $\mathbf{z}_a^{(i)}$ and $\mathbf{z}_b^{(i)}$ form a positive pair, $s(\cdot, \cdot)$ denotes the cosine similarity function, and $\tau_{\rm e}$ is a temperature hyperparameter. To ensure a balanced, bidirectional alignment between the two modalities, we define a symmetric InfoNCE loss:
\begin{equation}
    \mathcal{L}_{\text{sym}}(\mathbf{Z}_a, \mathbf{Z}_b) = \frac{1}{2} \left[ \mathcal{L}_{\text{InfoN}}(\mathbf{Z}_a, \mathbf{Z}_b) + \mathcal{L}_{\text{InfoN}}(\mathbf{Z}_b, \mathbf{Z}_a) \right].
\end{equation}

For the image modality, we directly employ the symmetric contrastive learning objective between the extracted image feature $\bm{v}^{\text{I}}$ and the CSI feature $\mathbf{z}^{\text{C}}$.  The loss is formulated as:
\begin{equation}
    \mathcal{L}^{\text{I}} = \mathcal{L}_{\text{sym}}(\bm{v}^{\text{I}}, \mathbf{z}^{\text{C}}).
\end{equation}

For the LiDAR modality, we incorporate both the original LiDAR feature $\bm{v}^{\rm L}$ and the two diffusion-augmented features $\tilde{\bm{v}}^{\rm L(1)}$, $\tilde{\bm{v}}^{\rm L(2)}$ into the CL framework. The overall objective consists of two components: a contrastive loss $\mathcal{L}^{\rm L}_{\text{contra}}$ and a diffusion training loss $\mathcal{L}_{\text{diff}}$. Specifically, we formulate a triplet contrastive loss to enforce consistency between three pairs: the original LiDAR features and CSI features, and the two augmented LiDAR features and CSI features:}
\begin{align}
\nonumber
\label{lidarloss}
\mathcal{L}_{\text{contra}} = &\lambda_1 \cdot \mathcal{L}_{\text{sym}}(\bm{v}^{\rm L}, \mathbf{z}^{\text{C}}) + \lambda_2 \cdot \mathcal{L}_{\text{sym}}(\tilde{\bm{v}}^{\rm L(1)}, \mathbf{z}^{\text{C}}) \\
&+ \lambda_3 \cdot \mathcal{L}_{\text{sym}}(\tilde{\bm{v}}^{\rm L(2)}, \mathbf{z}^{\text{C}}).
\end{align}

In parallel to the above contrastive objective, we incorporate a diffusion training loss to optimize the sampler $\epsilon_\theta(\cdot)$. This component follows the standard DDPM objective, minimizing the mean-squared error between the predicted and actual noise:
\begin{equation}
    \mathcal{L}_{\text{diff}} = \mathbb{E}_{m,\bm{v}^{\rm L},\epsilon} \left[ \| \epsilon - \epsilon_\theta(\tilde{\bm{v}}^{\rm L}_m, m) \|^2 \right],
\end{equation}
with $m \sim \mathcal{U}\{1, \dots, T\}$ and $\epsilon \sim \mathcal{N}(0,\mathbf{I})$.
The complete training objective for WiFo-M$^2$-LiDAR thus combines both losses with a balancing weight:
{
\begin{equation}
    \mathcal{L}^{\text{L}} = \mathcal{L}_{\text{contra}} + \lambda_{\text{diff}} \cdot \mathcal{L}_{\text{diff}},
\end{equation}}where $\lambda_{\text{diff}} = 0.3$ in our implementation. 
% This multi-objective framework simultaneously learns robust cross-modal alignments and high-quality feature augmentations, enabling the model to capture both invariant semantic representations and diverse feature variations.

\vspace{-0.2em}
\section{Experiments}
% In this section, we first introduce the pre-training datasets for WiFo-M$^2$ and the testing datasets used to evaluate its ability to enhance wireless communications.
In this section, we begin by detailing the construction of the pre-training and testing datasets.
Next, we provide a detailed description of the network architecture and training hyper-parameters. Finally, we comprehensively evaluate and analyze the performance of WiFo-M$^2$.

% \vspace{-1em}
\subsection{Datasets}
\label{Datasets}

\begin{table*}[t]
\centering
\renewcommand\arraystretch{1.15}  % 设置行高
\footnotesize
\caption{{Overview of Pre-training Datasets.}}
\label{tab: preDataset}
\begin{tabular}{ccccccccc}
\toprule
Dataset Type & Data Source & Scenario & \makecell[c]{Carrier \\ Frequency (GHz)} & Dataset ID  & \makecell[c]{BS Antenna \\ Configuration} & \makecell[c]{Bandwidth \\ (GHz)} & \makecell[c]{Number of \\ Subcarriers} & Samples  \\ 
\midrule

% SynthSoM section
\multirow{13}{*}{\makecell[c]{Pre-training Dataset}} 
& \multirow{7}{*}{M$^3$SC \cite{M3SC}} 
& \multirow{7}{*}{\makecell[c]{Intersection \\ traffic}}  & \multirow{7}{*}{$28$} 
&  I1 & $32\times1$ & $0.02$  & $512$ & $7,048$ \\  
\cline{5-9}
& & & & I2  & $64\times1$  & $1$  & $1024$ & $13,888$ \\  
\cline{5-9}
& & & & I3  & $16\times8$  & $0.8$  & $512$ & $7,088$ \\  
\cline{5-9}
& & & & I4  & $16\times8$ & $0.4$  & $512$ & $7,168$ \\  
\cline{5-9}
& & & & I5  & $8\times8$  & $0.8$  & $1024$ & $14,048$ \\  
\cline{5-9}
& & & & I6  & $8\times4$ & $1$  & $256$ & $3,484$ \\  
% \midrule
\cline{2-9}
% M3SC section
& \multirow{5}{*}{SynthSoM \cite{synthsom}} 
& \multirow{5}{*}{\makecell[c]{Dense \\ building}}  & \multirow{5}{*}{$4.95$} &
B1  & $16\times16$ & $0.05$ & $256$  & $9,080$\\  
\cline{5-9}
& & & & B2  & $16\times8$ & $0.05$ & $512$ & $16,784$ \\  
\cline{5-9}
& & & & B3  & $8\times8$ & $0.05$ & $512$ & $14,816$ \\  
\cline{5-9}
& & & & B4 & $8\times4$ & $0.02$ & $512$ & $18,824$ \\  
\cline{5-9}
& & & & B5 & $4\times4$  & $0.1$ & $1024$ & $38,064$ \\ 
\bottomrule
\end{tabular}
% \vspace{-0.6em}
\end{table*}

\begin{table*}[t]
\centering
\footnotesize
\renewcommand\arraystretch{1.6}  % 设置行高
\caption{{Overview of Testing Datasets.}}
\label{tab: Dataset2}
\begin{tabular}{ccccccccc}
\toprule
Dataset Type & Data Source & Scenario & \makecell[c]{Carrier \\ Frequency (GHz)} & Dataset ID  & \makecell[c]{BS Antenna \\ Configuration} & \makecell[c]{Bandwidth \\ (GHz)} & \makecell[c]{Number of \\ Subcarriers} & Samples  \\ 
\midrule
% Testing Set (zero shot)

\multirow{7}{*}{\makecell[c]{Testing Dataset \\ (\textbf{Seen} Scenarios)}} 
& \multirow{7}{*}{\makecell[c]{SynthSoM \cite{synthsom}\\M$^3$SC \cite{M3SC}}} 
& \multirow{7}{*}{\makecell[c]{Intersection\\ traffic \\\&\\Dense\\ building}} & \multirow{7}{*}{$28$, $4.95$} 
& L1 & $16\times8$ & \makecell[c]{$0.8$ \\$0.05$} & $512$ & $809$ \\  
\cline{5-9}
&  &   & & L2 & $4\times4$ & $0.1$ & $1024$ & $547$ \\  
\cline{5-9}
& & & & L3 & $16\times16$ & $0.05$ & $256$ & $300$ \\  
\cline{5-9}
& & & & L4 & $8\times4$ & $1$ & $256$ & $229$ \\  
\cline{5-9}
& & & & L5 & \makecell[c]{$32\times1$ \\ $8\times4$} & \makecell[c]{$0.02$ \\$0.02$} & $512$ & $725$ \\  
\cline{5-9}
& & & & L6 & \makecell[c]{$64\times1$ \\ $8\times8$} & \makecell[c]{$1$ \\ $0.8$} & $64$ & $404$ \\  
\cline{5-9}
& & & & L7 & $8\times8$ & $0.05$ & $512$ & $437$ \\  
\midrule
\multirow{3}{*}{\makecell[c]{Testing Dataset \\ (\textbf{Unseen} Scenarios)}} 
& \multirow{1}{*}{SynthSoM-Twin \cite{synthsom-t}} 
& \multirow{1}{*}{Campus} & $5.9$
& C1 & $64\times1$ & $0.02$ & $64$ & $480$ \\  
\cline{2-9}
& \makecell[c]{DeepSense-6G \cite{DS6G} \\ (Measured data)}
& Scenario 31  & $60$ & S1 & $16\times1$ & $-$ & $-$ & $387$ \\  
\cline{2-9}
& ViWi \cite{alrabeiah2020viwi} 
& dist\_cam & $60$ & V1 & $64\times1$ & $0.5$ & $256$ & $1302$ \\  
\bottomrule
\end{tabular}
\vspace{-0.2em}
\end{table*}

We construct a multi-modal pre-training dataset that includes LiDAR, RGB images, and CSI data. The CSI data is collected from the links between the BS and the users. In all datasets, the users are equipped with a single antenna, while the BS employs either a UPA or ULA configuration, with adjacent antenna spacing set to half the wavelength at the central frequency. The data is sourced from the intersection traffic scenario in the M$^3$SC dataset~\cite{M3SC} and the dense building scenario in the SynthSoM dataset~\cite{synthsom}.  Various communication system configurations are considered, covering a wide range of subcarrier numbers, bandwidths, and antenna sizes. The detailed simulation configurations of the pre-training datasets are presented in Table \ref{tab: preDataset}.

We also construct a testing set, with detailed simulation configurations shown in Table \ref{tab: Dataset2}. The testing set consists of two parts: $7$ datasets for \textbf{seen} scenarios and $3$ datasets for \textbf{unseen} scenarios.
First, to validate the in-distribution performance of WiFo-M$^2$, we construct datasets L1 to L7, which are selected from the same links as the pre-training datasets. Among these, datasets L1, L5, and L6 are formed by merging several links with the same number of antennas and subcarriers but smaller data volumes. To test the cross-scenario generalization ability of WiFo-M$^2$, we construct datasets C1, S1, and V1 based on three new scenarios not included in the pre-training datasets, namely, the campus scenario in the SynthSoM-Twin dataset~\cite{synthsom-t}, the ``dist\_cam'' scenario in the ViWi dataset \cite{alrabeiah2020viwi}, and the Scenario 31 in the DeepSense-6G dataset \cite{DS6G}, respectively. Notably, these three datasets operate at entirely \textbf{new} carrier frequencies, and their images and LiDAR point clouds exhibit visual appearances and spatial layouts \textbf{unseen} in the pre-training data. This pronounced gap between training and evaluation domains highlights the strong universality and generalization capability of WiFo‑M$^2$. Fig.~\ref{fig:dataset} provides an overview of the multi-modal environment sensory data examples from all the aforementioned scenarios.
\begin{figure}
    \centering
    \includegraphics[width=0.97\linewidth]{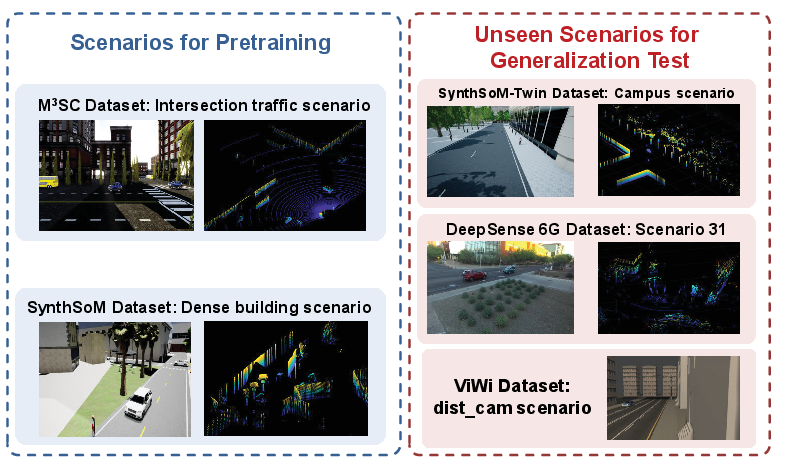}
    \caption{Overview of multi-modal sensory data from all datasets.}
    \label{fig:dataset}
    \vspace{-0.3em}
\end{figure}

 % \vspace{-0.8em}
\subsection{Network and Pre-training Settings}
The WiFo-M$^2$-LiDAR is trained for $100$ epochs using a batch size of $1024$. The optimization employs the AdamW optimizer with a learning rate of $1 \times 10^{-4}$ and a weight decay of $5\times 10^{-4}$. A cosine annealing learning rate scheduler with a warm-up phase is utilized: the learning rate linearly increases from zero to the target learning rate over the first $10$ epochs, then decays following a cosine schedule to zero by the final epoch. The contrastive learning objective is optimized with an initial temperature parameter $\tau_{\rm e}=0.1$ and the output feature dimension $d$ set to $512$. We set the loss weights in Eq.~\eqref{lidarloss} to $\lambda_1 = 1$, $\lambda_2 = 0.25$, and $\lambda_3 = 0.25$. The WiFo-M$^2$-Img shares the same hyperparameter settings as WiFo-M$^2$-LiDAR, except that the learning rate is set to $5 \times 10^{-4}$. 

 % \vspace{-1em}
\subsection{Performance Evaluation}
\label{performance}
In this work, we select {\color{black}five representative PHY actions: beam prediction (BP), pilot-based channel estimation (CE), channel interpolation (CI), channel prediction (CP), and wireless localization (WL), which together span the whole CSI pipeline from acquisition to utilization and underpin not only reliable network connectivity but also precise spatial localization}, providing a comprehensive test bed for our framework. {\color{black}In practical networks, varying operating frequencies (e.g., mmWave and sub-6 GHz) and scenario properties necessitate distinct algorithmic paradigms, such as compressive sensing for sparse channels or dense denoising for rich multipath environments. Therefore, we deliberately implement multiple distinct baselines for the same PHY action to demonstrate that WiFo-M$^2$ serves as a universal, paradigm-agnostic performance booster capable of enhancing heterogeneous algorithmic frameworks.}

The testing dataset is further divided into a training set and a validation set with a ratio of $4:1$, referred to as the T-T dataset and the T-V dataset, respectively. Note that in all subsequent experiments, the existing schemes are trained on T-T dataset and their performance is evaluated on T-V dataset. During this process, the weight of the WiFo-M$^2$ remains \textbf{frozen}, and only the existing schemes and the adapter module are \textbf{trained}.
\subsubsection{\textbf{PHY Action 1-Beam Prediction}}
\begin{itemize}
    \item[]
 \textbf{\textit{Problem Description}}: BP refers to directly predicting the optimal beamforming vector $\mathbf{f}^{*} \in \mathcal{F}$ from a pre-defined codebook $\mathcal{F}=\{ \mathbf{f}_{q}  \}_{q=1}^{Q}$ using multi-modal sensing. Let $\mathbf{h}\in \mathbb{C}^{ 1 \times N}$ be the CSI of a certain subcarrier. The optimal vector $\mathbf{f}^{*}=\mathop{\arg\max}_{m \in \{1,2,\cdots,Q\}}|\mathbf{h}\mathbf{f}_m|^2$ results in the maximum effective channel gain and should be chosen for transmission. The BP action can be formulated as: $\mathbf{f}^{*}=\mathcal{Q}_{\rm BP}(\mathbf{M})$,
     where $\mathcal{Q}_{\rm BP}(\cdot)$ represents a certain BP scheme and $\mathbf{M}$ is the multi‑modal sensory data.

    \quad Unlike the other three PHY actions, BP entirely relies on multi-modal sensing. Therefore, the WiFo-M$^2$ model is directly connected to a $2$-layer multilayer perceptron (MLP) network to output the optimal beam index. During fine-tuning on the T-T dataset, only the MLP is trained. Under this setting, the BP performance intuitively reflects the effectiveness of the OOB channel-aware features.
\end{itemize}

\begin{itemize}
    \item[]
    \textbf{\textit{Existing BP schemes}}: We selected two existing BP schemes, namely Vision-BP \cite{vision-beam} and MM-BP \cite{mm-beam}. Vision-BP \cite{vision-beam} detects transmitters with the YOLOv3 and then feeds the bounding-box centers to a NN to select the beam index. MM-BP \cite{mm-beam} is a multi‑modal fusion based scheme that extracts features from images, radar, and LiDAR.
\end{itemize}

\begin{figure*}
    \centering
    \includegraphics[width=1\linewidth]{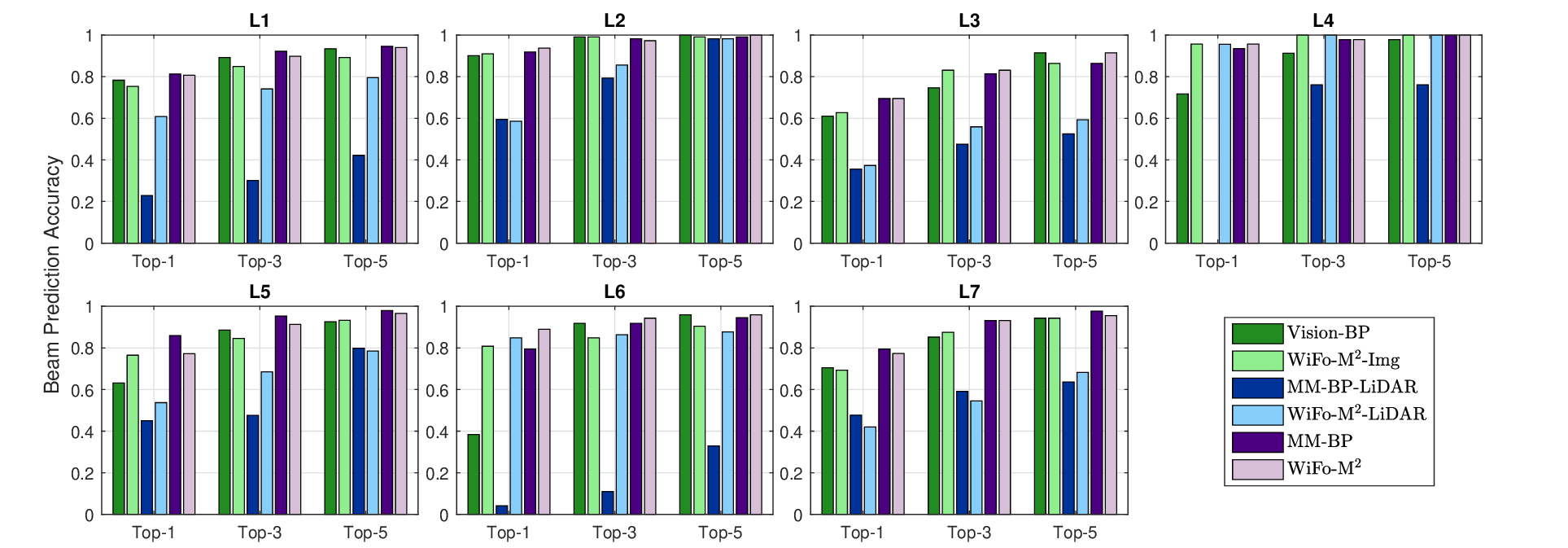}
    \caption{The Top-$k$ accuracies ($k\in\{1,3,5\}$) of the WiFo-M$^2$-based BP scheme, Vision-BP scheme, and MM-BP scheme.}
    \label{fig:BP}
    % \vspace{-0.2em}
\end{figure*}

\begin{itemize}
    \item[]
    \textit{\textbf{Experiment Settings}}:
    % Experiment Settings: We adopted the same training hyperparameters as \cite{vision-beam} and \cite{mm-beam} to train the Vision-BP and  MM-BP schemes. Specifically, for the MM-BP scheme, the batch size is set to $8$, with an initial learning rate of $1\times10^{-3}$. MM-BP-LiDAR is obtained by isolating only the LiDAR branch of the original MM-BP model, yielding a single-modality LiDAR baseline for beam prediction. The learning rate is reduced by a factor of $0.1$ at epochs $40$, $80$, and $100$. For the Vision-BP scheme, the batch size is $32$, and the initial learning rate is $1\times10^{-4}$, which decays by a factor of $0.1$ at epochs $4$ and $8$. 
    The training hyperparameters for Vision-BP and MM-BP schemes are kept identical to those used in \cite{vision-beam} and \cite{mm-beam}.
    The proposed WiFo-M$^2$-based BP scheme is trained with a batch size of $32$ and an initial learning rate of $2\times10^{-4}$. 
\end{itemize} 

\begin{itemize}
    \item[]
    \textit{\textbf{WiFo-M$^2$'s Performance}}: We evaluate the performance of BP using the top-$k$ accuracy metric. This metric measures the percentage of test samples for which the ground-truth optimal beam is included in the model's $k$ highest-ranked predictions. In our experiments, we report results for $k = 1, 3, 5$ to comprehensively assess the prediction capability.
    
    \quad As shown in Fig.~\ref{fig:BP}, across all $7$ datasets, WiFo-M$^2$ models exhibit highly competitive prediction accuracy even though the backbones are frozen. The WiFo-M$^2$-Img scheme attains noticeably higher Top-1 prediction accuracy than Vision-BP scheme on L4, L5 and L6, whereas on the remaining four datasets its performance is only slightly lower than that of Vision-BP. This demonstrates that the WiFo-M$^2$-Img backbone acquires channel-aware knowledge extracted during pre-training. The effect is more pronounced on LiDAR-based BP: WiFo-M$^2$-LiDAR substantially surpasses the fully-trained MM-BP-LiDAR baseline on every dataset, which indicates that the pre-trained  WiFo-M$^2$-LiDAR captures spatial patterns that are highly predictive of the optimal beam.  When both modalities are available, WiFo-M$^2$ delivers virtually the same performance as the MM-BP scheme, with a mean Top-1 gap of only 2\%. In summary, the BP results confirm that the ContraSoM pre-training aligns visual and LiDAR features with the wireless channel domain. Consequently, high-quality beam selection can be achieved with minimal retraining overhead.
\end{itemize}

% \begin{table*}[!t]
% \setlength{\tabcolsep}{5pt} % 
% \renewcommand{\arraystretch}{1.0} % 调整行高
% \centering
% \caption{The Top-$k$ accuracies ($k\in\{1,3,5\}$) of the WiFo-M$^2$-based BP scheme, Vision-BP scheme, and MM-BP scheme.}
% \label{table_example}
% {\footnotesize
% \begin{tabular}{c | c | c | c | c | c | c | c}
% \toprule
% \textbf{\diagbox[width=25mm]{schemes}{Datasets}} & L1 & L2 & L3 & L4 & L5 & L6 & L7 \\ 
% \midrule
% Vision-BP \cite{vision-beam} & {\underline{$-6.75$} / \underline{$-7.92$}} & {$-6.56$ / $-7.76$} & {$-6.60$ / $-7.70$} & {\bm{$-6.78$} / \bm{$-8.01$}} & {$-4.88$ / $-4.38$} & {$-2.51$ / $-2.59$} & {$-6.27$ / $-7.01$} \\ 
% \midrule
% MM-BP \cite{mm-beam} & {\underline{$-5.17$} / \underline{$-6.11$}} & {$-5.05$ / $-6.02$} & {$-5.02$ / $-5.94$} & {\bm{$-5.27$} / \bm{$-6.32$}} & {{$-1.76$} / {$-0.52$}} &{$0.70$ / $-0.12$} & {$-4.60$ / $-4.49$} \\ 
% \midrule
% WiFo-M$^2$-I & {\bm{$-3.88$} / \bm{$-5.59$}} & {$-3.76$ / $-5.24$} & {$-3.81$ / $-5.39$} & {\underline{$-3.77$} / \underline{$-5.53$}} & {$-1.56$ / $-1.78$} & {$0$ / $-0.01$} & {$-3.72$ / $-4.99$}\\
% \midrule
% WiFo-M$^2$-L  & {\bm{$-3.88$} / \bm{$-5.59$}} & {$-3.76$ / $-5.24$} & {$-3.81$ / $-5.39$} & {\underline{$-3.77$} / \underline{$-5.53$}} & {$-1.56$ / $-1.78$} & {$0$ / $-0.01$} & {$-3.72$ / $-4.99$}\\

% \bottomrule
% \end{tabular}}
% \end{table*}

\subsubsection{\textbf{PHY Action 2-Channel Estimation}}

\begin{itemize}
    \item[]
    \textbf{\textit{Problem Description}}: CE action refers to the process in which the BS estimates the channel $\mathbf{h}$ at the pilot-placed subcarriers. We assume that the BS adopts the hybrid (analog/digital) structure  and is equipped with $N_{\rm{RF}}$ RF chains. The digital and analog precoding matrices are represented by $\mathbf{F}_{\rm{B}} \in \mathbb{C}^{N_{\rm{RF}} \times N_{\rm{RF}}}$ and $\mathbf{F}_{\rm{R}} \in \mathbb{C}^{N \times N_{\rm{RF}}}$ with $\Vert \mathbf{F}_{\rm{R}}\mathbf{F}_{\rm{B}} \Vert_{\rm{F}}^2=N_{\rm{RF}}$. $\mathbf{F}_{\rm{B}}$ is set as an identity matrix $\mathbf{I}_{N_{\rm{RF}}}$  during CE. The quantization bit number of the phase shifters adopted at BS is set to $B=3$. 
    Each element of $\mathbf{F}_{\rm{R}}$ is randomly chosen from the set $\mathcal{A}_{Q}=\bigl\{\,e^{j\frac{2\pi q}{Q}} \,\bigm|\,q=1,\dots,Q\bigr\}$, where $Q\triangleq2^B=8$.
    % $\{e^{\frac{j2\pi\times1}{Q}}, e^{\frac{j2\pi\times2}{Q}},\cdots,e^{\frac{j2\pi\times Q}{Q}}\}$, where $Q\triangleq2^B=8$.
    During CE, the single-antenna user transmits known pilots $\mathbf{x} \in \mathbb{C}^{1\times T}$ over $T$ successive time slots. The received signal sequence at BS $\mathbf{Y} \in \mathbb{C}^{N_{\rm RF}\times T}$ can then be expressed as:
    \begin{equation}
        \mathbf{Y} = \mathbf{F}_{\rm{B}}^{\rm H}\mathbf{F}_{\rm{R}}^{\rm H} \mathbf{h}^{\rm H} \mathbf{x} + \mathbf{F}_{\rm{B}}^{\rm H}\mathbf{F}_{\rm{R}}^{\rm H}\mathbf{N},
    \end{equation}
    where $\mathbf{N}\in \mathbb{C}^{N\times T}$ denotes the additive white Gaussian noise with i.i.d. entries of zero mean and variance $\sigma^2$.
    Upon receiving the pilots, BS employs a certain CE network to estimate the CSI: $\hat{\mathbf{h}} = \mathcal{Q}_{\text{CE}}\bigl(\mathbf{Y},\mathcal{G}(\mathbf{M}))$,
    where $\mathcal{Q}_{\rm CE}(\cdot)$ represents a certain CE scheme and $\mathcal{G}(\cdot)$ denotes the WiFo‑M$^2$ model.
\end{itemize}

\begin{itemize}
    \item[]
    \textbf{\textit{Existing CE schemes}}: We selected two existing CE schemes, namely CENN~\cite{ma2020sparse} and FCDAMP \cite{LAMP}. 
    % CENN~\cite{ma2020sparse} is a DL-based compressed sensing channel estimation scheme. Its core idea is to predict the beamspace channel amplitude, and then reconstruct the channel by selecting the dominant channel entries accordingly. FCDAMP~\cite{LAMP} is a CE scheme that combines a fully convolutional denoising network with learned approximate message passing. It extracts noise features through a noise estimation sub-network and recovers the channel via a denoising sub-network.
    CENN~\cite{ma2020sparse} predicts the beam-space channel magnitudes and reconstructs the channel from the dominant coefficients, whereas FCDAMP~\cite{LAMP} inserts a fully convolutional denoising network into a learned approximate message passing loop to iteratively remove noise and recover the channel.
\end{itemize}

\begin{itemize}
    \item[]
    \textbf{\textit{Experiment Settings}}: The number of time slots $T$ is set to $8$ for the CENN scheme and $16$ for the FCDAMP scheme, while the number of RF chains $N_{\rm RF}$ is $4$ for all test datasets and schemes.  
    The learning rate for all schemes is set to $1\times10^{-4}$, and the batch size is $32$.
    % , the weight decay is set to $5\times10^{-5}$
\end{itemize} 

\begin{itemize}
    \item[]
    \textbf{\textit{WiFo-M$^2$'s Performance}}: The normalized mean square error (NMSE) is used as the estimation performance metric given by ${\rm NMSE}   =	{{\Vert \hat{\mathbf{H}}-\mathbf{H} \Vert}_{\rm{F}}^2}/{{\Vert \mathbf{H} \Vert}_{\rm{F}}^2}$. We evaluate the channel estimation performance of CENN and FCDAMP schemes under varying SNR conditions. As shown in Fig.~\ref{fig:CE}, the WiFo-M$^2$-enhanced variants consistently outperform the original CENN and FCDAMP models across most datasets and SNRs, demonstrating the general utility of the OOB channel-aware features. Furthermore, multi-modal fusion (WiFo-M$^2$) typically yields the most robust and significant gains, with LiDAR-only and image-only enhancements providing comparable improvements in most scenarios. 
    
    {\color{black}\quad It is observed that the performance gains brought by WiFo-M$^2$ are relatively marginal on the L3 dataset. This limitation arises from the extreme undersampling condition. While L3 is configured with a massive $16\times16$ antenna array, it uses the same limited number of pilots as other datasets. Consequently, the baseline CE performance is fundamentally poor due to the lack of pilot measurements. Under such condition, the coarse-grained environmental prior provided by WiFo-M$^2$ is insufficient to fully compensate for the missing physical observations, resulting in reduced enhancement.}
% On the L3 dataset with a large $256$-antenna array and limited pilots, the performance gains are less pronounced and sometimes irregular. This aligns with the expectation that under extremely sparse measurements and high-resolution array requirements, the coarse prior provided by multi-modal sensing becomes less effective in refining the estimation.
\end{itemize}

\begin{figure*}
    \centering
    \includegraphics[width=1.03\linewidth]{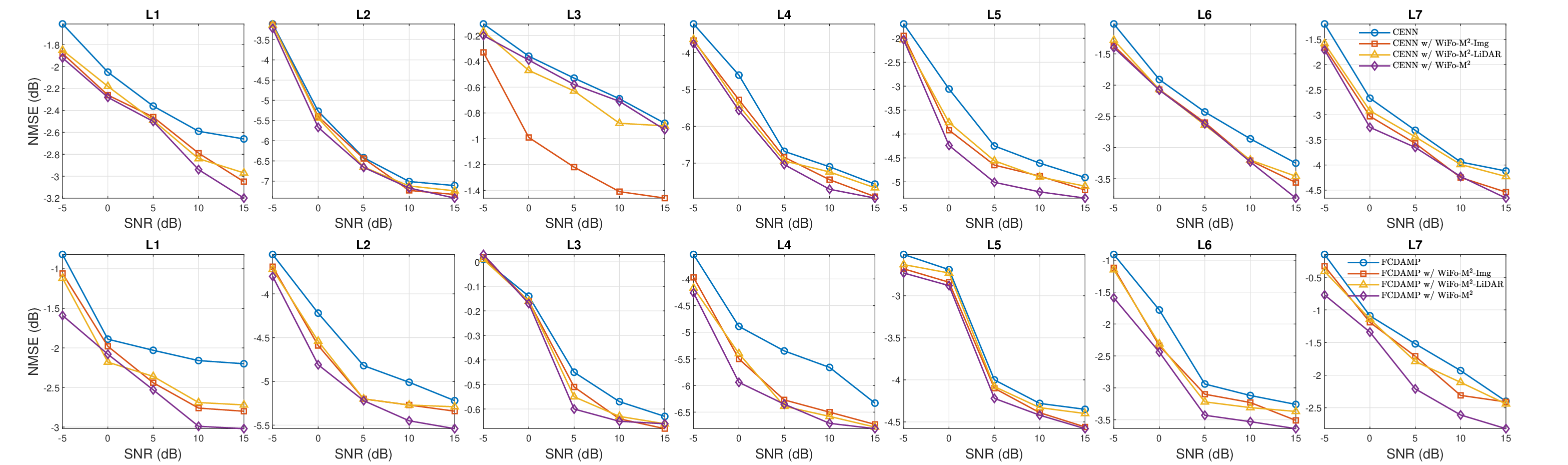}
    \caption{CE Performance of CENN and FCDAMP with WiFo-M$^2$, WiFo-M$^2$-Img, and WiFo-M$^2$-LiDAR.}
    \label{fig:CE}
    \vspace{-0.5em}
\end{figure*}

\subsubsection{\textbf{PHY Action 3-Channel Interpolation}}

\begin{itemize}
    \item[]
    \textbf{\textit{Problem Description}}: During the CE stage, pilots are placed only on a subset of subcarriers, and a subset of BS's antenna elements is activated\footnote{In CI action, we assume that a subset of antennas are provided with coarse channel estimates, while the CI action are required to recover the full CSI across all antennas.  Although this setup does not reflect a practical system configuration, it serves as an effective testbed for evaluating the evaluated CI schemes' ability to perform spatial super‑resolution reconstruction.}. Upon receiving pilots, BS estimates the channel at pilot positions using the Least Squares (LS) algorithm. All the available channel estimates form the low-resolution (LR) channel observation, denoted by $\hat{\mathbf{H}}^{\rm LR}\in \mathbb{C}^{K_{\rm p} \times N_{\rm p}}$. Specifically, if pilots are placed every $k$ subcarrier and every $x$ antenna element, the LR observation can be expressed as:
    \begin{equation}
        \hat{\mathbf{H}}^{\rm LR}[i,j] = \mathbf{H}[1:k:K,1:x:N]+\mathbf{N}[i,j],
    \end{equation}
     where $\mathbf{N}$ denotes the estimation noise. Then, BS utilizes a certain CI scheme to recover the high-resolution (HR) channel image $\hat{\mathbf{H}}\in \mathbb{C}^{K \times N}$. The entire process can be formulated as: $\hat{\mathbf{H}}=\mathcal{Q}_{\rm CI}(\hat{\mathbf{H}}^{\rm LR},\mathcal{G}(\mathbf{M}))$,
    where $\mathcal{Q}_{\rm CI}(\cdot)$ represents a certain CI scheme.
    \end{itemize}

\begin{figure*}
    \centering
    \includegraphics[width=1\linewidth]{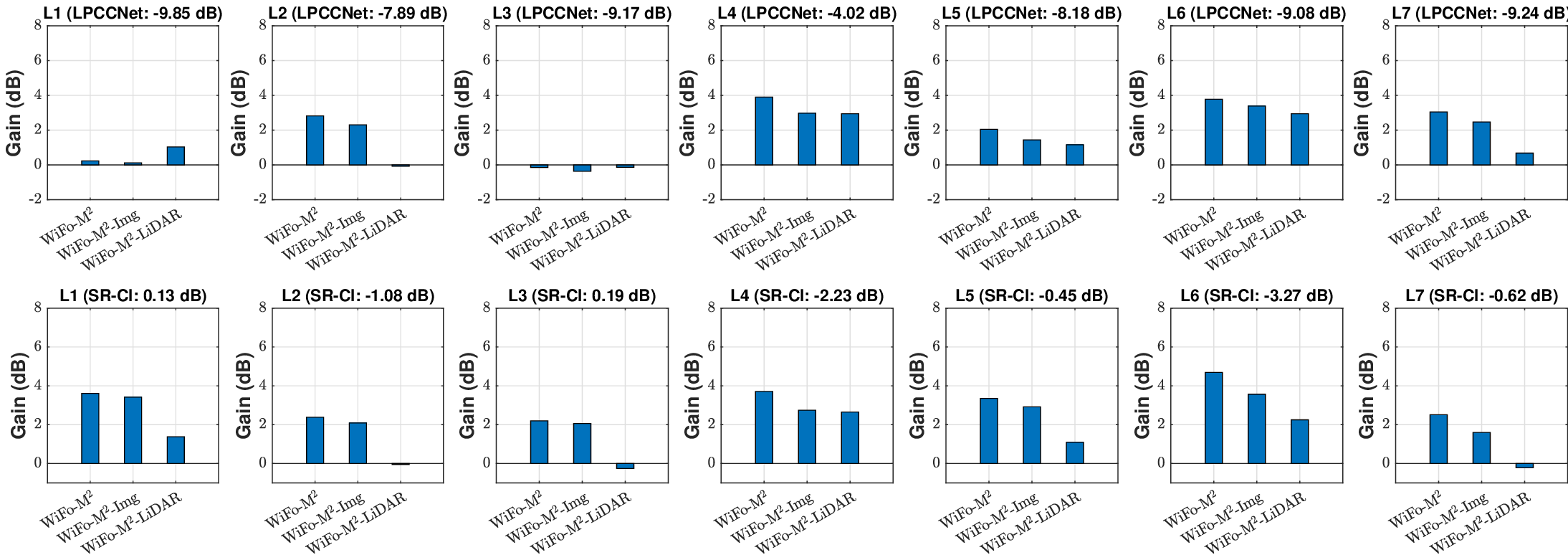}
    \caption{Performance gain brought by WiFo-M$^2$, WiFo-M$^2$-Img, and WiFo-M$^2$-LiDAR on  LPCCNet and SR-CI.}
    \label{fig:CI}
    % \vspace{-0.2em}
\end{figure*}

\begin{itemize}
    \item[]
    \textbf{\textit{Existing CI schemes}}: We selected two existing CI schemes, namely SR-CI~\cite{IRS} and LPCCNet \cite{lpcc}. SR-CI~\cite{IRS} obtains a low-precision and low-resolution channel matrix via LS and then feeds the linearly interpolated result into a super-resolution network to enhance the accuracy of CI. LPCCNet~\cite{lpcc} is a Laplacian pyramid‑based channel completion network that progressively reconstructs the high‑resolution channel image from a low‑resolution input through a stepwise up‑sampling architecture.
\end{itemize}

\begin{itemize}
    \item[]
    \textbf{\textit{Experiment Settings}}: We assume that the pilot placement interval in the subcarrier dimension and the sampling ratios for antenna dimensions are set to $k=K/8$ and $x=N/8$, respectively. That is, coarse channel estimates are available on only $1/8$ of the subcarriers and $1/8$ of the antennas, respectively. For network hyper-parameters, the SR-CI scheme uses a learning rate of $1\times10^{-2}$, the LPCCNet scheme uses $1\times10^{-4}$. The batch size for both schemes is $32$.
     % with a weight decay of $1\times10^{-5}$
     %  with no weight decay
\end{itemize} 

\begin{itemize}
    \item[]
    \textbf{\textit{WiFo-M$^2$'s Performance}}: The performance gains of WiFo-M$^2$ models over LPCCNet and SR-CI schemes are illustrated in Fig.~\ref{fig:CI}. Among the three variants, WiFo-M$^2$ consistently achieves the most stable and significant improvements across different datasets. This demonstrates the effectiveness of multi-modal fusion in enhancing CI accuracy. {\color{black}Note that on the L1 and L3 datasets, the WiFo-M$^2$ models provide relatively marginal performance gains or even introduce a slight degradation. This phenomenon is primarily attributed to the complex and dynamic channel variations present in these datasets across the spatial and frequency domains. In environments characterized by severe multipath fading, the macroscopic environmental priors extracted from sensing modalities become less effective at precisely tracking the small-scale fading dynamics. Consequently, the supplementary features provided by WiFo-M$^2$ offer limited assistance to the CI algorithms under these highly complex conditions.} Moreover, the performance gains brought by WiFo-M$^2$ are especially pronounced on the L4 and L6 datasets. L4 employs a BS with a small antenna array, while L6 is configured with relatively fewer subcarriers. Under these reduced spatial- or frequency-resolution settings, the OOB features provided by WiFo-M$^2$ are just sufficient to recover the correspondingly lower-dimensional CSI, leading to the sharp performance boost.
\end{itemize}

% \begin{table*}[!t]
% \setlength{\tabcolsep}{5pt} % 
% \renewcommand{\arraystretch}{1.0} % 调整行高
% \centering
% \caption{The NMSE performance of the SR-CI and its variations enhanced by different WiFo-M$^2$ model variants on channel interpolation.}
% \label{table_example}
% {\small
% \begin{tabular}{c | c | c | c | c | c | c | c}
% \toprule
% \textbf{\diagbox[width=25mm]{schemes}{Datasets}} & L1 & L2 & L3 & L4 & L5 & L6 & L7 \\ 
% \midrule
% SR-CI \cite{IRS} & {\underline{$-6.75$} / \underline{$-7.92$}} & {$-6.56$ / $-7.76$} & {$-6.60$ / $-7.70$} & {\bm{$-6.78$} / \bm{$-8.01$}} & {$-4.88$} & {$-2.51$ } & { $-7.01$} \\ 
% \midrule
% D17 & {\underline{$-5.17$} / \underline{$-6.11$}} & {$-5.05$ } & {$-5.02$ / $-5.94$} & {\bm{$-5.27$} / \bm{$-6.32$}} & {{$-1.76$} / {$-0.52$}} &{$0.70$ } & {$-4.60$ } \\ 
% \midrule
% D18 & {\bm{$-3.88$} / \bm{$-5.59$}} & {$-3.76$ } & {$-3.81$ } & {\underline{$-3.77$} / \underline{$-5.53$}} & {$-1.56$ / $-1.78$} & {$0$ / $-0.01$} & {$-3.72$ }\\
% \midrule
% D18 & {\bm{$-3.88$} / \bm{$-5.59$}} & {$-3.76$ } & {$-3.81$} & {\underline{$-3.77$} / \underline{$-5.53$}} & {$-1.56$ / $-1.78$} & {$0$ / $-0.01$} & {$-3.72$ }\\

% \bottomrule
% \end{tabular}}
% \end{table*}

\subsubsection{\textbf{PHY Action 4-Channel Prediction}}

\begin{itemize}
    \item[ ]
    \textbf{\textit{Problem Description}}: In frequency division duplex (FDD) systems, where uplink-downlink reciprocity does not hold, the BS relies on uplink channel feedback to obtain downlink CSI. However, the feedback overhead scales linearly with the number of antennas and devices, making it infeasible for massive MIMO systems. Given that uplink and downlink transmissions experience the same physical environment, it is reasonable to utilize the uplink CSI measured at the BS to predict the downlink CSI. Similarly to \cite{frequency_CP_ref1,frequency_CP_ref2}, the first half of the subcarriers is considered as the uplink channel, while the remaining half is treated as the downlink channel. Let $K_{\rm in}$ be the number of subcarriers in the first half and $K_{\rm out}$ represent the number of subcarriers in the second half. Then, the uplink channel $\mathbf{H}_{\rm d}$ and downlink channel $\mathbf{H}_{\rm u}$ can be expressed as:
    \begin{equation}
        \mathbf{H}_{\rm d} = \mathbf{H}[1:K_{\rm in},:], \mathbf{H}_{\rm u} = \mathbf{H}[K_{\rm in}+1:K_{\rm in}+K_{\rm out},:].
    \end{equation}
    The CP action can be formulated as: $\mathbf{H}_{\rm u}=\mathcal{Q}_{\rm CP}({\mathbf{H}}_{\rm d},\mathcal{G}(\mathbf{M}))$,
     where $\mathcal{Q}_{\rm CP}(\cdot)$ represents a certain CP scheme.
\end{itemize}

\begin{itemize}
    \item[]
    \textbf{\textit{Existing CP schemes}}: We selected two CP schemes, namely WiFo \cite{wifo} and Transformer-based scheme \cite{Dai_Trans}. WiFo \cite{wifo} is a wireless FM designed for time/frequency-domain CP, which is pretrained via several self-supervised training tasks. The transformer-based scheme \cite{Dai_Trans} is a parallel CP scheme.
\end{itemize}

\begin{table*}[!t]
\setlength{\tabcolsep}{5pt} % 
\renewcommand{\arraystretch}{0.5} % 调整行高
\centering
\caption{CP performance of the WiFo and Transformer-based models, along with their variations enhanced by different WiFo-M$^2$ models. The left and right sides of the slash represent the WiFo and Transformer-based schemes, respectively.}
\label{table_CP}
{\footnotesize
\begin{tabular}{c | c | c | c | c | c | c | c}
\toprule
\textbf{\diagbox[width=25mm]{Schemes}{Datasets}} & L1 & L2 & L3 & L4 & L5 & L6 & L7 \\ 
\midrule
CP baselines & {$-5.88$\big/$-0.60$} & {$-2.41$\big/$-0.91$} & {$-4.51$\big/$-1.38$} & {$-0.93$\big/$-2.74$} & {$-2.36$\big/$0$} & {$-4.39$\big/$-1.15$} & {$-2.78$\big/$-1.22$} \\ 
\midrule
\makecell[c]{CP baselines \\ + WiFo-M$^2$-Img} &  {$-5.84$\big/$-0.70$} & {\bm{$-2.75$}\big/$0$} & {$-5.19$\big/$-2.05$} & {$-1.15$\big/$-3.25$} & {$-2.83$\big/$1.21$} & {$-4.73$\big/$-1.13$} & {$-3.05$\big/$-1.21$}\\
\midrule
\makecell[c]{CP baselines \\ + WiFo-M$^2$-LiDAR} &  {$-5.92$\big/$-0.69$} & {$-2.68$\big/\bm{$-1.43$}} & {{$-5.07$}\big/$-1.90$} & {$-1.09$\big/$-3.12$} & {{$-3.41$}\big/$0$} & {{$-5.04$}\big/\bm{$-1.32$}} & {$-2.94$\big/$-1.07$}\\
\midrule
\makecell[c]{CP baselines \\ + WiFo-M$^2$} & {\bm{$-6.02$}\big/\bm{$-0.76$}} & {$-2.12$\big/$-1.12$} & {\bm{$-5.27$}\big/\bm{$-2.25$}} &  {\bm{$-1.27$}\big/\bm{$-3.64$}} & {\bm{$-3.61$}\big/$0$} &{\bm{$-5.10$}\big/$-1.28$} & {\bm{$-3.30$}\big/{$-1.22$}} \\ 
\bottomrule
\end{tabular}}
% \vspace{-0.5em}
\end{table*}

\begin{itemize}
    \item[]
    \textbf{\textit{Experiment Settings}}: The learning rates for the Transformer-based scheme and WiFo are set to $1\times 10^{-3}$ and $1\times 10^{-5}$, respectively. The batch size is set to $32$. The network input $\mathbf{H}_{\rm d}$ is assumed to be noisy CSI estimates with an NMSE of $-20$dB.
    % The weight decay for Transformer-based scheme is $1\times 10^{-5}$, while that for WiFo is $1\times 10^{-2}$. 
\end{itemize} 

\begin{itemize}
    \item[]
    \textbf{\textit{WiFo-M$^2$'s Performance}}: The performance enhancements provided by WiFo-M$^2$ models on top of baseline WiFo and Transformer-based schemes are summarized in Table \ref{table_CP}. The NMSE is also used as performance metric. The WiFo-M$^2$ model leveraging multi-modality demonstrates its superiority by delivering highest gains for both CP baselines across most datasets, underscoring the benefit of cross-modal feature fusion. Consistent with observations in CI action, performance gains are particularly significant on datasets L6. Notably, WiFo-M$^2$'s effectiveness is inherently bounded by the convergence of the baseline scheme itself. For instance, on dataset L5, the Transformer-based baseline fails to achieve downlink CSI prediction, and correspondingly, WiFo-M$^2$ provides no enhancement. This indicates that the supplementary features cannot compensate for fundamental algorithmic limitations in the baseline model. 

    {\color{black}\quad While multi-modal fusion generally achieves the highest gains, we observe slight instability on certain datasets (e.g., L2 and L6), where the multi-modal variant performs comparably to, or even slightly worse than, the single-modal counterparts. This instability stems from the straightforward feature concatenation strategy currently employed for multi-modal fusion. It suggests that simple concatenation may occasionally introduce feature redundancy, indicating that the joint exploitation of multi-modal features has not yet been fully optimized.}
\end{itemize}

{\color{black}
\subsubsection{\textbf{PHY Action 5-Wireless Localization}}

\begin{figure}
    \centering
    \includegraphics[width=1\linewidth]{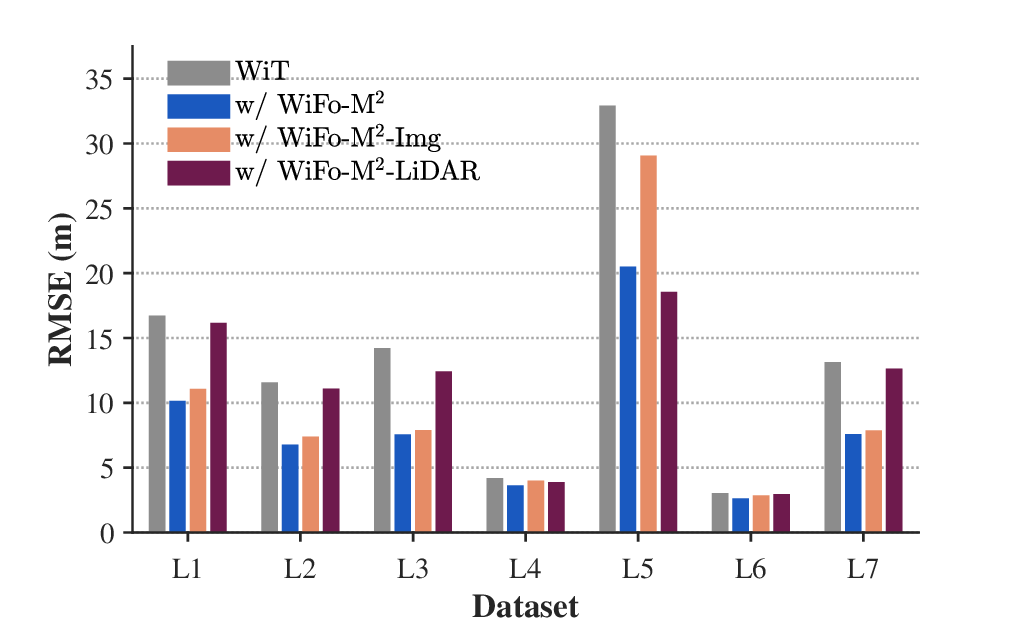}
    \caption{{\color{black}Localization performance of WiT with WiFo-M$^2$, WiFo-M$^2$-Img, and WiFo-M$^2$-LiDAR.}}
    \label{fig:WL}
    % \vspace{-0.2em}
\end{figure}

\begin{itemize}
    \item[ ]
    \textbf{\textit{Problem Description}}: CSI inherently captures the rich multi-carrier and multi-antenna spatial signatures of the surrounding physical environment. Consequently, it can be effectively utilized as a highly distinctive spatial fingerprint for accurate positioning. We formulate the WL task as a direct mapping from the estimated CSI observations to the physical location of the UE. Specifically, the network input is defined as the complete CSI $\mathbf{H} \in \mathbb{C}^{K \times N}$. The output of the network is the estimated 2D coordinate $\hat{\mathbf{p}} = [\hat{x}, \hat{y}]^T \in \mathbb{R}^{2}$ of the UE. The WL action can be formulated as $\hat{\mathbf{p}} = \mathcal{Q}_{\text{WL}}(\mathbf{H})$, where $\mathcal{Q}_{\text{WL}}(\cdot)$ represents a certain WL network.
\end{itemize}

\begin{itemize}
    \item[]
    \textbf{\textit{Existing WL schemes}}: We selected WiT \cite{WiT}, a Transformer-based framework that leverages multi-head self-attention to capture global dependencies across subcarriers and antennas. By efficiently distilling spatial features from multi-path environments, it serves as a robust baseline for wireless localization. 
\end{itemize}

\begin{itemize}
    \item[]
    \textbf{\textit{Experiment Settings}}: During training, the batch size is fixed at $32$. The learning rates are chosen as $2\times10^{-5}$. Additionally, we assume the input $\mathbf{H}$ is noisy CSI estimates with a $-20\text{ dB}$ NMSE.
\end{itemize}

\begin{itemize}
    \item[]
    \textbf{\textit{WiFo-M$^2$'s Performance}}: The root mean square error (RMSE) is adopted as the metric to measure the deviation between the estimated coordinates and the true coordinates, defined as: ${\rm RMSE}   =	\sqrt{\frac{1}{n}\sum_{i=1}^{n}(\mathbf{p}-\hat{\mathbf{p}})^2}$. The localization performance gains achieved by incorporating the proposed WiFo-M$^2$ models into WiT are illustrated in Fig.~\ref{fig:WL}. 
    
    \quad As shown in Fig.~\ref{fig:WL}, the multi-modal WiFo-M$^2$ framework consistently delivers the most stable and significant RMSE reductions across diverse datasets. Remarkably, empowered by the environmental priors provided by WiFo-M$^2$, the localization errors on the vast majority of datasets are effectively suppressed to below $10$ m. Note that the localization errors across all schemes remain relatively large on the L5 dataset. This is primarily attributed to the harsh NLoS blockages in the dense building scenario and the minimum system bandwidth configured for this dataset. Physically, narrow bandwidth constrains the multi-path delay resolution, causing adjacent paths to overlap in the frequency domain and impairing CSI-based geometric inference. Nonetheless, incorporating WiFo-M$^2$ substantially reduces the RMSE from $32.92$ m to $20.51$ m. 
    % Moreover, the single-modal WiFo-M$^2$-LiDAR achieves the best performance on L5 with an RMSE of $18.56$ m. This implies that the precise 3D geometry captured by sparse point clouds provides critical absolute structural constraints, effectively calibrating the unreliable signal-based positioning features under narrow-band constraints.
\end{itemize}

}

% \vspace{-1em}
\subsection{Ablation Study}
% In this subsection, we investigate the influence of various specialized designs in both the ContraSoM pre-training strategy and WiFo-M$^2$ model, including temporal feature extrapolation (denoted by TFE) and diffusion-based LiDAR feature augmentation (denoted by DA). To highlight the crucial role of the ContraSoM strategy, we additionally evaluate WiFo-M$^{2}$-Img whose weights are replaced by ImageNet-pre-trained ones, denoted as WiFo-M$^{2}$-Img w/ INet Weight.

% Due to page limits, we present ablation results for several representative PHY actions, as shown in Table \ref{table_aba1} and Table \ref{table_aba2}. First, removing the temporal feature extrapolation function in WiFo-M$^2$ leads to a performance drop, confirming that extrapolating fine-grained future features effectively bridges the sparsity of sensory data. The NMSE results reported are averaged over the L1–L7 datasets. The ablation study results demonstrate the effectiveness of each proposed component. Third, removing the DA design from WiFo-M$^{2}$-LiDAR causes a performance decline, validating that the proposed feature‑space diffusion process enhances LiDAR representation learning.
{\color{black}
As detailed in Table \ref{table_aba1} and Table \ref{table_aba2}, we conduct a comprehensive ablation study to validate the individual contributions of the temporal feature extrapolation (TFE), pixel-level image augmentation (PA), and diffusion-based LiDAR feature augmentation (DA) modules. First, removing the TFE module in both the image and LiDAR sub-models results in a performance drop across almost all datasets. This confirms that extrapolating fine-grained future features is essential to effectively bridge the temporal sparsity of sensory data during inter-frame perception gaps. Second, eliminating the PA module in WiFo-M$^2$-Img leads to a noticeable degradation across downstream actions. Finally, disabling the DA module in WiFo-M$^2$-LiDAR causes a decline in overall performance. This validates that the proposed feature-space diffusion process successfully enriches sparse point cloud representations, compelling the network to learn robust and channel-aware geometric structures.}

\begin{table*}[!t]
% 策略3：进一步缩小列间距以换取内部文字不被缩得太小
\setlength{\tabcolsep}{3pt} 
% 增加行高，因为现在每个单元格有两行文字，需要空间避免拥挤
\renewcommand{\arraystretch}{1.2} 
\centering
\caption{{\color{black}Ablation of PA and TFE modules on WiFo-M$^2$-Img.}}
\label{table_aba1}
% 策略2：使用 resizebox 强制整体宽度不超过文本总宽 (\textwidth)
\resizebox{\textwidth}{!}{%
% === 开启全局蓝色字体与蓝色表格线 ===
\color{black}
\begin{tabular}{c | c | c | c | c | c | c | c}
\toprule
\textbf{\diagbox[width=28mm]{Schemes}{Datasets}} & L1 & L2 & L3 & L4 & L5 & L6 & L7 \\ 
\midrule
\makecell[c]{PHY action baselines \\ \makecell[c]{(CE/CI \\ CP/WL)}} 
& \makecell[c]{$-2.36 / -9.85$ \\ $-5.88 / 16.72$} 
& \makecell[c]{$-6.42 / -7.89$ \\ $-2.41 / 11.58$} 
& \makecell[c]{$-0.53$ / \bm{$-9.17$} \\ $-4.51 / 14.22$} 
& \makecell[c]{$-6.68 / -4.02$ \\ $-0.93 / 4.20$} 
& \makecell[c]{$-4.25 / -8.18$ \\ $-2.36 / 32.92$} 
& \makecell[c]{$-2.43 / -9.08$ \\ $-4.39 / 3.04$} 
& \makecell[c]{$-3.31 / -9.24$ \\ $-2.78 / 13.14$} \\ 
\midrule
\makecell[c]{Baselines + \\ WiFo-M$^2$-Img} 
& \makecell[c]{\bm{$-2.46$} / \bm{$-9.97$} \\ $-5.84$ / \bm{$11.08$}} 
& \makecell[c]{\bm{$-6.44$} / $-10.19$ \\ \bm{$-2.75 / 7.40$}} 
& \makecell[c]{$-0.72$ / $-8.80$ \\ \bm{$-5.19 / 7.90$}} 
& \makecell[c]{\bm{$-6.84 / -6.99$} \\ \bm{$-1.15 / 4.01$}} 
& \makecell[c]{\bm{$-4.65 / -9.62$} \\ \bm{$-2.83 / 29.06$}} 
& \makecell[c]{\bm{$-2.60 / -12.47$} \\ \bm{$-4.73 / 2.87$}} 
& \makecell[c]{\bm{$-3.57 / -11.71$} \\ \bm{$-3.05 / 7.87$}} \\
\midrule
\makecell[c]{Baselines + \\ WiFo-M$^2$-Img w/o TFE} 
& \makecell[c]{$-2.32 / -9.34$ \\ \bm{$-5.89$} / $11.55$} 
& \makecell[c]{$-4.79$ / \bm{$-10.61$} \\ $-2.69 / 7.99$} 
& \makecell[c]{\bm{$-0.87$} / $-8.32$ \\ $-5.08 / 8.10$} 
& \makecell[c]{$-4.69 / -6.35$ \\ $-1.05 / 4.25$} 
& \makecell[c]{$-3.32 / -9.27$ \\ $-2.71 / 32.47$} 
& \makecell[c]{$-2.19 / -10.63$ \\ $-4.71 / 3.92$} 
& \makecell[c]{$-2.53 / -9.19$ \\ $-2.96 / 8.36$} \\
\midrule
\makecell[c]{Baselines + \\ WiFo-M$^2$-Img w/o PA} 
& \makecell[c]{$-2.39 / -9.82$ \\ $-5.84 / 18.14$} 
& \makecell[c]{$-6.43 / -9.62$ \\ $-2.68 / 10.56$} 
& \makecell[c]{$-1.05 / -8.52$ \\ $-4.98 / 9.21$} 
& \makecell[c]{$-6.57 / -6.91$ \\ $-0.95 / 4.34$} 
& \makecell[c]{$-4.29 / -9.06$ \\ $-2.56 / 31.92$} 
& \makecell[c]{$-2.56 / -9.52$ \\ $-4.67 / 2.91$} 
& \makecell[c]{$-3.40 / -10.97$ \\ $-3.04 / 11.81$} \\ 
\bottomrule
\end{tabular}%
} % 结束 resizebox
\end{table*}

\begin{table*}[!t]
% 策略3：进一步缩小列间距以换取内部文字不被缩得太小
\setlength{\tabcolsep}{3pt} 
% 增加行高，因为现在每格有两行文字，需要足够空间避免上下重叠
\renewcommand{\arraystretch}{1.2} 
\centering
\caption{{\color{black}Ablation of DA and TFE modules on WiFo-M$^2$-LiDAR.}}
\label{table_aba2}
% 策略2：使用 resizebox 强制整体宽度不超过文本总宽 (\textwidth)
\resizebox{\textwidth}{!}{%
% === 开启全局蓝色字体与蓝色表格线 ===
\color{black}
% \arrayrulecolor{black}
\begin{tabular}{c | c | c | c | c | c | c | c}
\toprule
\textbf{\diagbox[width=28mm]{Schemes}{Datasets}} & L1 & L2 & L3 & L4 & L5 & L6 & L7 \\ 
\midrule
\makecell[c]{PHY action baselines \\ \makecell[c]{(CE/CI \\ CP/WL)}} 
& \makecell[c]{$-2.36 / -9.85$ \\ $-5.88 / 16.72$} 
& \makecell[c]{$-6.42 / -7.89$ \\ $-2.41 / 11.58$} 
& \makecell[c]{$-0.53$ / \bm{$-9.17$} \\ $-4.51 / 14.22$} 
& \makecell[c]{$-6.68 / -4.02$ \\ $-0.93 / 4.20$} 
& \makecell[c]{$-4.25 / -8.18$ \\ $-2.36 / 32.92$} 
& \makecell[c]{$-2.43 / -9.08$ \\ $-4.61 / 3.04$} 
& \makecell[c]{$-3.31 / -9.24$ \\ $-2.78 / 13.14$} \\ 
\midrule
\makecell[c]{Baselines + \\ WiFo-M$^2$-LiDAR} 
& \makecell[c]{\bm{$-2.49 / -10.89$} \\ \bm{$-5.92 / 16.17$}} 
& \makecell[c]{\bm{$-6.65 / -7.81$} \\ \bm{$-2.68$} / {$11.10$}} 
& \makecell[c]{\bm{$-0.63$} / $-9.03$ \\ \bm{$-5.07 / 12.43$}} 
& \makecell[c]{\bm{$-6.94 / -6.96$} \\ \bm{$-1.09 / 3.88$}} 
& \makecell[c]{$-4.56$ / \bm{$-9.34$} \\ \bm{$-3.41 / 18.56$}} 
& \makecell[c]{\bm{$-2.64 / -12.02$} \\ \bm{$-5.05 / 2.96$}} 
& \makecell[c]{\bm{$-3.44$} / $-9.92$ \\ \bm{$-2.94$} / \bm{$12.64$}}\\
\midrule
\makecell[c]{Baselines + \\ WiFo-M$^2$-LiDAR w/o TFE} 
& \makecell[c]{$-2.30 / -9.75$ \\ $-5.84 / 17.35$} 
& \makecell[c]{$-4.95 / -7.79$ \\ $-2.47 / 11.31$} 
& \makecell[c]{$-0.47 / -8.78$ \\ $-4.60 / 12.94$} 
& \makecell[c]{$-4.97 / -5.85$ \\ $-0.95 / 4.24$} 
& \makecell[c]{$-3.27 / -8.02$ \\ $-2.85 / 30.73$} 
& \makecell[c]{$-2.43 / -9.64$ \\ $-4.99 / 3.71$} 
& \makecell[c]{$-2.65 / -9.35$ \\ $-2.92 / 12.39$}\\
\midrule
\makecell[c]{Baselines + \\ WiFo-M$^2$-LiDAR w/o DA} 
& \makecell[c]{$-2.25 / -10.40$ \\ $-5.81 / 17.04$} 
& \makecell[c]{$-6.64 / -7.26$ \\ $-2.51$ / \bm{$10.92$}} 
& \makecell[c]{$-0.55 / -8.61$ \\ $-5.04 / 12.48$} 
& \makecell[c]{$-6.69 / -6.56$ \\ $-0.95 / 3.40$} 
& \makecell[c]{\bm{$-4.66$} / $-7.83$ \\ $-2.96 / 19.64$} 
& \makecell[c]{$-2.47 / -10.92$ \\ $-4.73 / 3.22$} 
& \makecell[c]{$-3.39$ / \bm{$-10.08$} \\ $-2.88$ / {$12.74$}} \\ 
\bottomrule
\end{tabular}%
} % 结束 resizebox
\end{table*}

{\color{black}Furthermore, to isolate the source of the performance gains and validate the semantic effectiveness of the extracted features, we conduct an additional iso-dimensional ablation study, as presented in Table \ref{ablation2}. First, we evaluate the impact of the pre-training strategy by replacing the ContraSoM pre-trained weights with generic ImageNet-pretrained weights (denoted as w/ INet Weight). While generic INet weights can occasionally provide moderate improvements by capturing basic visual semantics (e.g., vehicles and buildings) as coarse spatial priors, their lack of domain specificity leads to inconsistent and sometimes detrimental results. For instance, on LPCCNet, ContraSoM delivers a pronounced $1.89$ dB NMSE gain, whereas INet weights actually degrade performance below the baseline (from $-7.74$ dB to $-7.55$ dB). This highlights the critical advantage of explicitly learning channel-aware features via ContraSoM pretraining over relying on general-purpose visual representations.

Second, to rigorously validate the efficacy of the WiFo-M$^2$'s features and ensure that the performance gains are not merely an artifact of increased model capacity, we decouple the impact of structural expansion from the actual semantic value of the features. When integrating the features extracted by WiFo-M$^2$ into the downstream PHY networks, the additional parameters inherently increase the model's degrees of freedom. To test this, we replace the WiFo-M$^2$'s features with uninformative tensors of the identical dimension: all-ones vectors (denoted as Weight-1), all-zeros vectors (Weight-0), and random Gaussian noise (Weight-R). As shown in Table \ref{ablation2}, injecting these uninformative variants yields only marginal fluctuations or even noticeable performance degradations compared to the baselines. For instance, the localization RMSE under uninformative weights fails to improve upon the WiT baseline ($16.47$ m). Although integrating uninformative tensors like Weight-0 occasionally provides a slight performance gain (e.g., reducing the NMSE of WiFo from $-3.08$ dB to $-3.31$ dB), such improvements are inferior to the enhancements delivered by WiFo-M$^2$. This outcome conclusively illustrates that the superiority of WiFo-M$^2$ models stems directly from their ability to extract critical channel-aware spatial and geometric priors from multi-modal sensing, rather than relying on parameter expansion.
}

\begin{table}[!t]
\setlength{\tabcolsep}{6pt} 
\renewcommand{\arraystretch}{0.7} % 调整行高
\centering
\caption{{\color{black}Iso-dimensional ablation study on feature effectiveness.}}
\label{ablation2}

% 使用 resizebox 限制表格宽度不超过当前列宽
\resizebox{\columnwidth}{!}{% <-- 注意这里的百分号，防止产生多余的空格
\begin{tabular}{c | c || c || c || c}
\toprule
 & CE: FCDAMP  & CI: LPCCNet & CP: WiFo & WL: WiT \\
 \midrule
WiFo-M$^{2}$-Img  & \bm{$-3.38$} dB & \bm{$-9.63$} dB & $-3.38$ dB & \color{black}{$12.93$ m} \\
\midrule
WiFo-M$^{2}$-LiDAR  & $-3.36$ dB & $-9.13$ dB & \bm{$-3.45$} dB & \color{black}{\bm{$12.35$} m} \\
 \midrule
\makecell[c]{WiFo-M$^{2}$-Img  \\ w/ INet Weight} & $-3.22$ dB & $-7.55$ dB & $-2.96$ dB & \color{black}{$13.00$ m}\\
\midrule
\makecell[c]{\color{black}{Fine-tune}  \\ \color{black}{w/ Weight-1}} & \color{black}{$-2.54$ dB} & \color{black}{$-5.72$ dB} & \color{black}{$-3.23$ dB} & \color{black}{$16.58$ m} \\
\midrule
\makecell[c]{\color{black}{Fine-tune}  \\ \color{black}{w/ Weight-0}} & \color{black}{$-2.41$ dB} & \color{black}{$-6.33$ dB} & \color{black}{$-3.31$ dB} & \color{black}{$16.22$ m} \\
\midrule
\makecell[c]{\color{black}{Fine-tune}  \\ \color{black}{w/ Weight-R}} & \color{black}{$-2.40$ dB} & \color{black}{$-5.94$ dB} & \color{black}{$-3.04$ dB} & \color{black}{$17.93$ m} \\
\midrule
{\color{black}Baseline} & {\color{black}$-3.14$ dB} & {\color{black}$-7.74$ dB} & {\color{black}$-3.08$ dB} & {\color{black}$16.47$ m} \\
\bottomrule
\end{tabular}% <-- 注意这里的百分号
}
\end{table}

% \vspace{-0.9em}
\subsection{Cross-Scenario Generalization}
In this section, we evaluate the cross-scenario generalization ability of WiFo-M$^2$ on the datasets C1, V1, and S1, which are not seen in the pre-training datasets.

\subsubsection{S1 Dataset}
The S1 dataset provides measured image and LiDAR point cloud data, as well as the receive power with beams. Due to the absence of CSI, only the performance of BP action is evaluated on S1 dataset. Note that BP action on S1 dataset is relatively straightforward due to the small 16-element antenna array. The Vision-BP baseline achieves Top-1/3/5 accuracies of 91.0\%, 98.7\%, and 100\%, while the MM-BP scheme reaches 92.2\%, 100\%, and 100\%. Notably, the WiFo-M$^2$-img model, with its pre-trained backbone frozen and only a final MLP layer fine-tuned, attains 91.0\%, 100\%, and 100\%, matching or surpassing the strong baselines. This demonstrates that the representations learned by WiFo-M$^2$ successfully generalize to the unseen S1 scenario.
% \begin{table}[!t]
% \setlength{\tabcolsep}{5pt} % 
% \renewcommand{\arraystretch}{1.0} % 调整行高
% \centering
% \caption{The Top-$k$ accuracies ($k\in\{1,3,5\}$) of the WiFo-M$^2$-based BP scheme, Vision-BP scheme, and MM-BP scheme on S1 dataset.}
% \label{table_example}
% {\footnotesize
% \begin{tabular}{c | c }
% \toprule
% \textbf{\diagbox[width=28mm]{Schemes}{Metric}}  &  Top-1/3/5 Accuracy  \\ 
% \midrule
% Vision-BP  & {$91.0\% / 98.7 \% / 100\%$}  \\ 
% \midrule
% MM-BP  & {$92.2\% / 100 \% / 100\%$} \\ 
% \midrule
% WiFo-M$^2$-based BP & {$91.0\% / 100 \% / 100\%$} \\
% \bottomrule
% \end{tabular}}
% \end{table}

\subsubsection{C1 Dataset}
Table \ref{table_C} summarizes the cross-scenario generalization performance of WiFo-M$^2$ variants when applied to CE, CI, CP, and WL actions on the unseen C1 dataset. 
As can be seen, the WiFo-M$^2$ model with multi-modality provides consistent and often the best performance gains across most schemes, achieving the lowest NMSE for both CE schemes and one CP scheme (WiFo). {\color{black}Similarly, WiT w/ WiFo-M$^2$ scheme achieves the lowest RMSE.} Importantly, the WiFo-M$^2$-Img consistently outperforms its ImageNet-pre-trained counterpart, confirming the effectiveness of the proposed ContraSoM strategy. BP results exhibit the same trend. Using only LiDAR, the MM‑BP-LiDAR baseline achieves top‑1/3/5 accuracy of $69.4\%$, $86.0\%$, and $86.0\%$, while WiFo-M$^2$-LiDAR attains $73.1\%$, $90.3\%$, and $99.5\%$. With image alone, the vision-BP baseline delivers $78.5\%$, $88.2\%$, and $90.3\%$, against $77.4\%$, $85.5\%$, and $89.2\%$ for WiFo-M$^{2}$-Img. Under multi-modal fusion, MM-BP scheme peaks at $88.7\%$, $99.5\%$, and $100\%$, whereas WiFo-M$^{2}$ achieves $83.9\%$, $91.4\%$, and $93.0\%$. 

\begin{table*}[!t]
\renewcommand{\arraystretch}{0.8} % 调整行高
\centering
\caption{Evaluation of the enhancement effects of WiFo-M$^2$ models on various PHY actions on C1 dataset. Values in bold indicate the best performance.}
\label{table_C}
% 使用 resizebox 将表格宽度限制在双栏的整体文本宽度内
\resizebox{\textwidth}{!}{
\begin{tabular}{c | c | c || c | c || c | c || c}
\toprule
\textbf{\diagbox[width=54mm]{Schemes}{PHY Actions}}  & CE: CENN & CE: FCDAMP & CI: LPCCNet & CI: SR-CI & CP: WiFo & CP: Transformer & {\color{black}WL: WiT}\\ 
\midrule
Baseline  & {$-4.74$} dB & {$-5.30$} dB & {$-16.03$} dB & {$-8.68$} dB & {$-10.74$} dB & {$-14.06$} dB & {\color{black}{$1.31$ m} } \\ %8
\midrule
\makecell[c]{Baseline  + WiFo-M$^2$-Img w/ INet Weight}  & {$-4.87$} dB & {$-5.38$} dB &  {$-16.73$} dB & {$-7.29$} dB &{$-11.25$} dB & {$-13.58$} dB & {\color{black}{$1.58$ m}}\\  %8
\midrule 
\makecell[c]{Baseline  + WiFo-M$^2$-Img w/ ContraSoM Weight}  & {$-4.96$} dB & {$-5.43$} dB & {$-17.21$} dB &  \bm{$-9.19$} dB & {$-11.58$} dB & {$-14.23$} dB & {\color{black}{$1.12$ m}}\\   %8
\midrule
\makecell[c]{Baseline  + WiFo-M$^2$-LiDAR}  & {$-4.77$} dB & {$-5.43$} dB & {$-17.24$} dB & {$-8.59$} dB & {$-11.45$} dB &  \bm{$-14.28$} dB & {\color{black}{$1.19$ m}}\\ %8
\midrule
\makecell[c]{Baseline  + WiFo-M$^2$}  & \bm{$-4.98$} dB & \bm{$-5.49$} dB &  \bm{$-17.29$} dB & {$-9.11$} dB & \bm{$-12.21$} dB & {$-14.26$} dB & \bm{{\color{black}{$0.94$ m} }}\\ %12
\bottomrule
\end{tabular}
}
\end{table*}

\begin{table*}[!t]
\renewcommand{\arraystretch}{0.7} % 调整行高
\centering
\caption{Evaluation of the enhancement effects of WiFo-M$^2$-Img model on various PHY actions on V1 dataset. Values in bold indicate the best performance.}
\label{table_V}
% 使用 resizebox 将表格宽度限制在双栏文本宽度内
\resizebox{\textwidth}{!}{
\begin{tabular}{c | c | c || c | c || c | c || c}
\toprule
\textbf{\diagbox[width=54mm]{Schemes}{PHY Actions}}  & CE: CENN & CE: FCDAMP & CI: LPCCNet & CI: SR-CI & CP: WiFo & CP: Transformer & {\color{black}WL: WiT}\\ 
\midrule
Baseline  & {$-3.09$ dB} & {$-3.17$ dB} & {$-8.81$ dB} & {$-3.85$ dB} & {$-5.23$ dB} & {$-$} & {\color{black}{$2.56$ m} } \\ 
\midrule
\makecell[c]{Baseline  + WiFo-M$^2$-Img w/ INet Weight}  & {$-2.33$ dB} & {$-4.28$ dB} &  {$-9.52$ dB} & {$-4.32$ dB} &{$-5.02$ dB}  & {$-$} & {\color{black}{$2.44$} m}\\ 
\midrule
\makecell[c]{Baseline  + WiFo-M$^2$-Img w/ ContraSoM Weight}  & \bm{$-3.41$} dB & \bm{$-5.17$} dB & \bm{$-10.46$} dB & \bm{$-4.35$} dB & \bm{$-5.55$} dB  & {$-$} & {\color{black}\bm{$2.33$} m}\\
\bottomrule
\end{tabular}
}
% \vspace{-0.5em}
\end{table*}

\subsubsection{V1 Dataset}
The V1 dataset provides only images and CSI data. Hence, we focus on the performance of WiFo-M$^2$-Img. In BP action, the WiFo-M$^2$-Img-based scheme achieves top‑1/3/5 accuracy of $67.4\%$, $91.2\%$, and $98.1\%$, outperforming Vision‑BP ($65.5\%$, $82.4\%$, $92.3\%$) and closely matching the performance of the MM‑BP scheme ($71.3\%$, $87.4\%$, $99.2\%$) which uses only its image branch. Table \ref{table_V} presents the enhancement effect of WiFo-M$^2$-Img on other baselines, demonstrating consistent performance improvements across all baselines, with the exception of the Transformer-based CP scheme, which failed to converge. In the same way, the weights derived from the ContraSoM strategy play a significant role. 

\begin{table}[!t]
\setlength{\tabcolsep}{5pt}      % 仅控制内部水平 padding
\renewcommand{\arraystretch}{0.9}
\centering
\caption{Additional fine-tunable parameters introduced by applying
WiFo-M$^2$ to different PHY actions.}
\label{table_parametes}
{\footnotesize
%
% ─── 第一段：5 列 ────────────────────────────────────────────────
\begin{tabular}{p{1.8cm}|cccc}
\toprule
{PHY Actions}            & CE: CENN & CE: FCDAMP & CI: LPCCNet & BP \\ % ← 若无第 4 个可删
\midrule
Parameters (M)   &  \makecell[c]{0.13\\0.26} & \makecell[c]{0.09\\0.18} & \makecell[c]{0.035\\0.051} & \makecell[c]{0.067\\0.13} \\
\bottomrule
\end{tabular}

\vspace{3pt}  % 段间距，可按需求调

% ─── 第二段：4 列，第一列宽度与上段一致，列间距拉大 ─────────────
\begin{tabular}{p{1.8cm}|cccc}
\toprule
{PHY Actions}   & CI: SR-CI & CP: WiFo & CP: Transformer & {\color{black}WL: WiT} \\
\midrule
Parameters (M) & \makecell[c]{0.017\\0.034} & \makecell[c]{0.26\\0.26}  & \makecell[c]{0.026\\0.055}   & {\color{black}\makecell[c]{0.26\\0.26}} \\
\bottomrule
\end{tabular}}
% \vspace{-0.4em}
\end{table}
% \subsection{Performance Analysis of Model's Size}

% \vspace{-0.7em}
{\color{black}\subsection{Robustness of WiFo-M$^2$ with Unannotated Sensory Data}
In practical deployment scenarios, the receiver may frequently be occluded by dynamic obstacles or fall outside the sensors' FoV, resulting in missing receiver labeling. To evaluate the robustness of WiFo-M$^2$ under such severe occlusions, we specifically construct a test subset comprising samples where the receiver is unannotated in both the LiDAR and image modalities. This experiment aims to investigate whether the WiFo-M$^2$ can still extract effective OOB channel-aware features and provide performance gains by relying solely on the captured background environmental context and global spatial priors.

\begin{figure}
    \centering
    \includegraphics[width=1\linewidth]{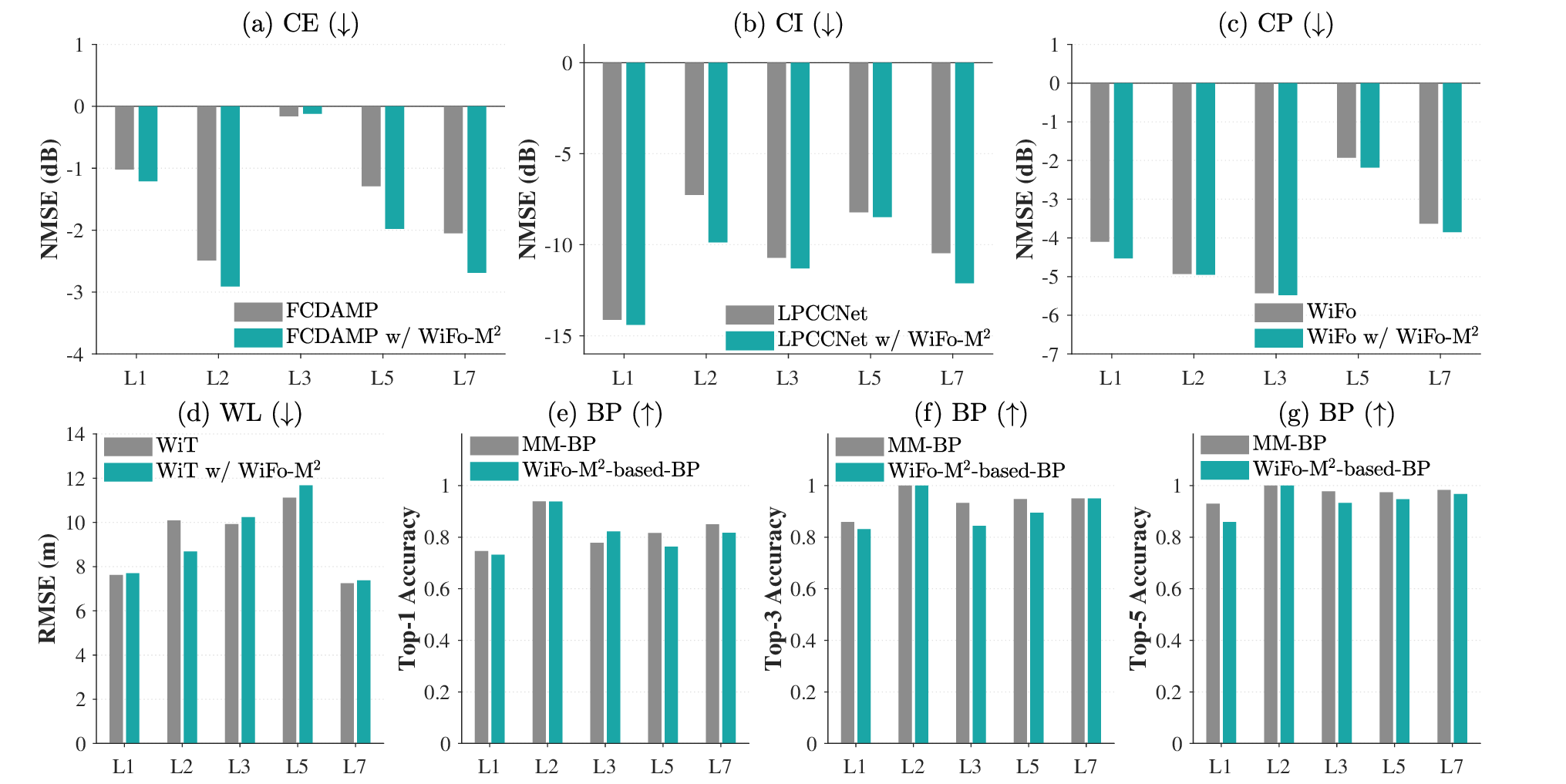}
    \caption{{\color{black}Performance enhancements achieved by WiFo-M$^2$ for CE, CI, CP, WL, and BP using unannotated sensory data.}}
    \label{fig:annoa}
    % \vspace{-0.7em}
\end{figure}

This evaluation utilizes datasets L1, L2, L3, L5, and L7, which contain valid samples lacking receiver labeling in both sensory modalities. As shown in Fig.~\ref{fig:annoa}, for CE, CI, and CP actions, WiFo-M$^2$ still delivers consistent gains. This indicates that even without explicit receiver coordinates, WiFo-M$^2$ successfully extracts generalized representations of the ambient electromagnetic environment. These static background contexts serve as robust spatial priors for multipath components, effectively improving channel reconstruction actions. Conversely, for WL and BP, the WiFo-M$^2$-enhanced schemes show marginal improvements or slight degradations compared to baselines. This divergence stems from distinct task dependencies: WL and BP are highly user-centric actions relying highly on absolute coordinates and angular directions. Without explicit receiver labeling serving as geometric anchors, it is difficult for the model to capture the critical directional cues required for these localized tasks, as generalized background context alone is insufficient.
}

\subsection{Computational and Deployment Efficiency}
To assess the practical deployment overhead of WiFo-M$^2$, we evaluate its model complexity and inference latency. All experiments are conducted on a server equipped with dual 14-core Intel Xeon E5-2680 v4 CPUs and an NVIDIA GeForce RTX 4090 GPU, processing inputs with a batch size of $1$ to simulate real-time operation. WiFo-M$^2$-LiDAR comprises only $0.07$M parameters with an inference latency of $3.31$ms, while WiFo-M$^2$-Img contains $9.62$M parameters with an inference latency of $6.04$ms. When multiple modalities are available, the total inference can be executed in (at most) the sum of their latencies, i.e., $9.35$ms, when processed serially. Even with this upper bound, the system stays well within a real-time budget: at $40$ fps a new multi-modal sample arrives every $25$ms, leaving $15.7$ms after inference to exploit the freshly produced features. During those $9.35$ms, the features predicted from the previous sample can still be used, which only requires the model’s output sequence to be extended to cover the $9.35$ms inference window.

As another practical metric, Table \ref{table_parametes} presents the additional fine-tunable parameters introduced when WiFo-M$^2$ is integrated into PHY actions. The parameter overhead remains extremely low, reaching at most $0.26$M. This is due to the effectiveness and superiority of the ContraSoM pre-training strategy, which allows the WiFo-M$^2$ backbone to remain frozen during adaptation and only requires the lightweight adapter to be fine-tuned together with the existing schemes. This further  highlights the practical deployability of WiFo-M$^2$.

% \vspace{-0.3em}
\section{Conclusions}
\label{conclusion}
In this paper, we have realized a framework where intelligent embodied network entities utilize multi-modal environment sensing to comprehensively refine their PHY actions in a plug-and-play manner. Central to our approach is a powerful foundation model, WiFo-M$^2$, which is capable of extracting fine-grained OOB channel-aware features from sparse historical sensing inputs. This is enabled by ContraSoM, a contrastive pre-training scheme that aligns environment sensing with wireless channels in a shared latent space.
% Empowered by large-scale pre-training, WiFo-M$^2$ is able to derive fine-grained OOB channel features from sparse historical sensing input, providing a transferable representation that plugs seamlessly into diverse transceiver modules.
% By providing a scalable and transferable representation, WiFo-M$^2$ fundamentally shifts the design paradigm from schemes bound to particular scenarios, system configurations, and PHY actions, toward a universal approach, thereby extensively extending the benefits of multi-modal sensing to a wider range of core PHY actions. 
To enhance representation robustness, we further incorporate modality-specific data augmentation techniques that apply stochastic pixel-space transformations for images and leverage a diffusion-based feature generation framework for LiDAR point clouds. 
Extensive experiments on typical PHY actions show that our proposed framework boosts network performance from on multiple fronts and generalizes well across diverse scenarios.

% \vspace{-0.7em}

\end{document}